\documentclass[aps,pra,amsmath,amssymb,superscriptaddress,showpacs,10pt,reprint
]{revtex4-1}

\usepackage{pbox}
\usepackage[dvips]{graphicx}
\usepackage[colorlinks=true]{hyperref}
\usepackage[normalem]{ulem}
\usepackage{xcolor}
\hypersetup{pdfpagemode = UseNone}
\usepackage{multirow}
\usepackage{afterpage}

\usepackage{braket}
\usepackage{dsfont}
\usepackage[rightcaption]{sidecap}  	
\sidecaptionvpos{figure}{t} 
\usepackage{enumitem}

\newcommand{\omegaC}{{\omega_\textrm{}}}

\newcommand{\betaC}{{\beta_\textsf{C}}}
\newcommand{\betaQ}{{\beta_\textsf{Q}}}

\newcommand{\fQ}{\textsf{Q}}
\newcommand{\fD}{\textsf{D}}
\newcommand{\fC}{\textsf{C}}

\newcommand{\fQC}{\textsf{QC}}
\newcommand{\fQDC}{\textsf{QDC}}

\newcommand{\nC}{{n_\textsf{C}}}
\newcommand{\nQ}{{n_\textsf{Q}}}

\newcommand{\mC}{{m_\textsf{C}}}
\newcommand{\mQ}{{m_\textsf{Q}}}
\newcommand{\mD}{{m_\textsf{D}}}

\newcommand{\rhof}{\rho_\text{f}}

\newcommand{\pb}{{p_\text{b}}}

\newcommand{\ePR}{{\epsilon_\text{prep}}}
\newcommand{\eRD}{{\epsilon_\text{read}}}
\newcommand{\eFB}{{\epsilon_\text{feed}}}
\newcommand{\eDT}{{\epsilon_\text{meas}}}
\newcommand{\eDTx}[1]{{\epsilon_\text{meas}^{#1}}}

\newcommand{\EP}[1]{{\Sigma_{#1}}}
\newcommand{\ep}{{\sigma}}

\newcommand{\setup}[1]{{\textsf{#1}}}
\newcommand{\eg}{e.g.,~}
\newcommand{\ie}{i.e.,~}

\newcommand{\Tr}{\mathrm{Tr}}
\newcommand{\mean}[1]{{\langle {#1}\rangle}}

\newcommand{\etad}{\varepsilon}
\newcommand{\Pa}{P_\mathrm{a}}
\newcommand{\na}{n_{\text{a}}}

\newcommand{\vph}{\vphantom{\text{\Huge G}}}
\newcommand{\vphh}{\vphantom{\text{\huge G}}}

\newcommand{\rhoQCk}{\rhof^{\fQC,k}}
\newcommand{\rhoQC}{\rhof^\fQC}
\newcommand{\rhoD}{\rhof^\fD}
\newcommand{\rhoQCD}{\rhof^\fQDC}
\newcommand{\I}{I_{\text{f}}^{\fQC:\fD}}

\begin{document}
\title{Alternative experimental ways to access entropy production}

\author{Zheng Tan}
\thanks{These authors contributed equally}
\affiliation{Laboratoire Kastler Brossel, Coll\`ege de France, CNRS, ENS-Universit\'e PSL, Sorbonne Universit\'{e}, 11 place Marcelin Berthelot, F-75231 Paris, France}
\author{Patrice A. Camati}
\thanks{These authors contributed equally}
\affiliation{Universit\'e Grenoble Alpes, CNRS, Grenoble INP, Institut N\'eel, 38000 Grenoble, France}
\author{Guillaume Cœuret Cauquil}
\affiliation{Laboratoire Kastler Brossel, Coll\`ege de France, CNRS, ENS-Universit\'e PSL, Sorbonne Universit\'{e}, 11 place Marcelin Berthelot, F-75231 Paris, France}
\author{Alexia Auff\`eves}
\affiliation{Universit\'e Grenoble Alpes, CNRS, Grenoble INP, Institut N\'eel, 38000 Grenoble, France}
\author{Igor Dotsenko}
\email{igor.dotsenko@lkb.ens.fr}
\affiliation{Laboratoire Kastler Brossel, Coll\`ege de France, CNRS, ENS-Universit\'e PSL, Sorbonne Universit\'{e}, 11 place Marcelin Berthelot, F-75231 Paris, France}
\date{\today}

\begin{abstract}


We theoretically derive and experimentally compare several different ways to access entropy production in a quantum process under feedback control. We focus on a bipartite quantum system realizing an autonomous Maxwell's demon scheme reported by Najera-Santos \textit{et al}.~[Phys.~Rev.~Research 2, 032025(R) (2020)], where information encoded in a demon is consumed to transfer heat from a cold qubit to a hot cavity. By measuring individual quantum trajectories of the joint demon-cavity-qubit system, we compute the entropy production with six distinct expressions
derived from different approaches to the system description and its evolution. Each method uses a specific set of trajectories and data processing. Our results provide a unified view on the various meanings of irreversibility in quantum systems and pave the way to the measurement of entropy production beyond thermal frameworks.

\end{abstract}
\maketitle

\section{Introduction}

Entropy production (EP) is a key physical concept that quantifies the irreversibility of a given process: the larger the EP, the more irreversible the process. It was born from very practical considerations, since irreversibility fundamentally limits the performance of heat engines and fridges~\cite{Bejanbook2016}.  Eventually it turned into {\it the} fundamental concept allowing to phrase the second law of thermodynamics (SLT): the EP of a physical process can never be negative. As a typical example, spontaneous heat flow from cold to hot bodies is forbidden by the SLT, as it would give rise to a negative
~EP.

Pioneering expressions of EP were established at the outset of macroscopic thermodynamics and generally applied to specific irreversible processes such as the thermalization of systems or, conversely, the driving out of their thermal equilibrium. Such processes involve heat dissipation into reservoirs of well-defined temperatures, therefore making heat and temperature two essential quantities to define EP. Later on, the ability to monitor and control the evolution of microscopic systems at the level of single realizations gave rise to the so-called stochastic EP~\cite{Esposito2009, Binderbook2018, Campisi2011}, that provided a renewed perspective on irreversibility. At this level of description, irreversibility results from random perturbations exerted on the  system dynamics by external reservoirs, thus preventing the external operator to rewind any protocol. In this view, EP fundamentally captures the lack of control over microscopic systems, a concept that broadens the notion of EP to a much wider range of situations. Moreover, stochastic thermodynamics is agnostic to the type of noise and reservoirs which cause irreversibility. Its conceptual tools can be adapted to any kind of random perturbation, holding the promise to quantify irreversibility of quantum nature, \eg stemming from decoherence or any source of quantum noise~\cite{Elouard2017,Landi2020}.

Microscopic systems undergoing feedback-controlled dynamics provide a first example of extension beyond open systems interacting with thermal environments. In such processes, information on the system's microstate is used to set its following evolution. In the past decades, it became possible to quantify the EP of these processes, evidencing a novel place for information within thermodynamics. Treated as a correlation between the controlled system and the memory of the feedback loop, information was shown to be an essential component of EP in experiments inspired in the Maxwell's demon paradox~\cite{Maxwellbook1975, Rex2017, Leff1990}.

From an experimental perspective, EP in its various acceptions was measured on a handful of platforms. Without feedback control, experiments at the ensemble average level have been performed in a nuclear magnetic resonance (NMR) setup~\cite{Batalhao2015}, in a micromechanical resonator~\cite{Brunelli2018} and in a Bose-Einstein condensate~\cite{Brunelli2018}. For feedback-controlled protocols, EP has been accessed at the average level in an NMR setup~\cite{Camati2016} and at the trajectory level with a superconducting circuit~\cite{Masuyama2018} and single-electron transistors \cite{Koski2015}. The latter case provides an example of an autonomous Maxwell's demon, where information is encoded on a quantum system and is never processed at the classical level. The device operated as a fridge, consuming information to transfer heat from a cold to a hot reservoir. In this spirit we have recently implemented a fully closed version of such a device where the cold and hot bodies, as well as the demon, are quantum systems evolving unitarily~\cite{Luis2020}. This situation is a minimalistic model of a closed, information-powered fridge. 

The ability of theoretically describing and experimentally realizing a wide range of irreversible processes involving an increasing number of parties has given rise to an equivalent variety of expressions of EP. This calls for the development of a unified perspective, serving as much as a consistency check for the various definitions and as a testbed for their respective sensitivity to measurement errors. This is the purpose of the present article, where we theoretically and experimentally study the EP of the model system recalled above~\cite{Luis2020}. Namely, we derive and compare six alternative methods to measure the entropy produced by this system which are chosen to cover and illustrate a large variety of equivalent approaches to characterize the EP. They differ by the way we analyze the system's state (ensemble average or quantum trajectories), theoretically describe the control (external or autonomous) and experimentally access the system evolution (single unitary evolution or a cyclic implementation incorporating the time reversal of the basic evolution). Each choice provides us with a different view onto the EP and its definition, allowing us to acquire a deeper understanding of the physical nature and the experimental meaning of EP obtained with different measurements. Despite being equivalent in the ideal case, these expressions show different sensitivities to experimental errors. This observation is confirmed by a thorough modelling of our experiment, providing a practical benchmark that can be used to adapt the measurement strategy of EP to a particular quantum system.

\section{Protocols and expressions}

\subsection{Review of general methods}

\begin{figure}[t]
\begin{centering}
\includegraphics[width=\columnwidth]{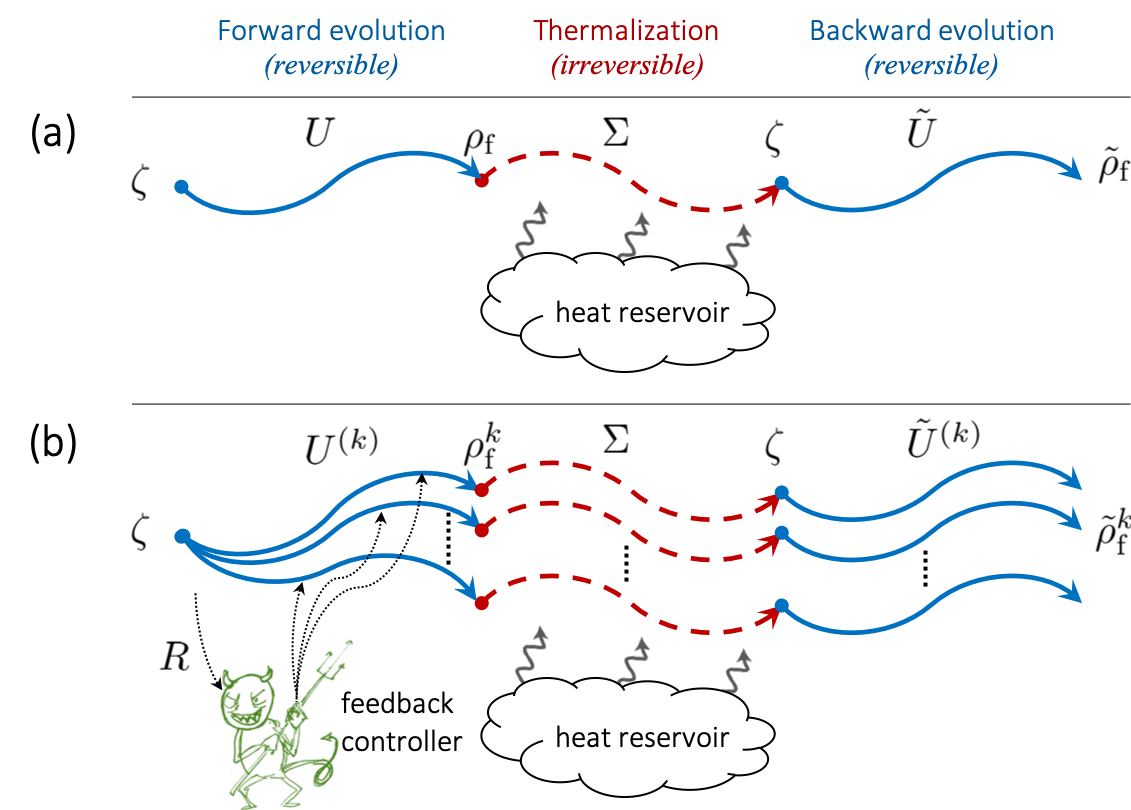}
\par\end{centering}
\caption{The concept of the forward and backward system's evolution used to access the entropy production of the thermalization process. (a) Starting in the equilibrium state $\zeta$, the system is unitarily driven out of equilibrium into state $\rho_\text{f}$. The irreversible thermalization with the external heat reservoir produces entropy $\EP{}$ by bringing the system back to $\zeta$. The backward evolution $\tilde U$ implements the time-reversed forward evolution. In the presence of the thermalization, the backward evolution cannot bring the system back into its initial state thus revealing the irreversibility. (b) The overall scheme can be extended to a feedback-controlled evolution, where the unitary $U^{(k)}$ depends on the result $k$ of a control measurement (readout $R$) of the system state by a feedback controller.}
\label{fig:entropy production}
\end{figure}

We first review the measures of irreversibility established within the so-called quantum Jarzynski's protocol \cite{An2015, Huber2008, Campisi2011, Batalhao2014, Heyl2012, Dorner2013, Mazzola2013}, schematically presented in Fig.~\ref{fig:entropy production}(a). After having thermalized with a heat reservoir at temperature $T$, a quantum system starts in the thermal equilibrium state $\zeta$ with inverse temperature $\beta = \left(k_{\text{B}}T\right)^{-1}$, also known as thermodynamic beta, where $k_{\text{B}}$ is the Boltzmann's constant. The system is first driven out of equilibrium through a unitary operation $U$, to the non-equilibrium state $\rho_\text{f}$ (here and in the following the subscript $\text{f}$ labels quantities at the end of the evolution). To be treated as a unitary, $U$ is assumed to be performed swiftly compared to the system relaxation. Then, the system relaxes back to the thermal state $\zeta$. This last step causes the whole process to be irreversible. The entropy production $\EP{}$ is proportional to the amount of heat $Q$ dissipated by the system along its thermalization, $\EP{} = \beta Q$. It is shown to equal the relative entropy $D(\rho_\text{f} ||\zeta)$, also known as quantum divergence, quantifying the non-negative distance between the two states. For states $\rho$ and $\sigma$, it is defined as $D(\rho||\sigma) = -\Tr[\rho\ln\sigma]-S(\rho)$, where $S(\rho)$ is the von Neumann entropy of state $\rho$~\cite{Vedral2002}. This provides a first intuitive flavour for the EP: the farther the system is brought away from equilibrium, the larger the entropy production.

Another meaning for the EP is acquired by the attempt to reverse the forward evolution $U$. For this purpose we complete the above protocol with the time-reversed unitary operation $\tilde U$. In general, unitary operations are considered as reversible: from an operational point of view this presupposes the ability to generate the backward evolution $U^\dagger = \tilde U$ (here and in the following the symbol $\sim$ denotes the backward quantities). In the absence of the intermediate thermalization, this backward evolution would bring the system back to its initial state. The presence of the intermediate irreversible thermalization is the reason why the process cannot be reversed. EP is shown to equal the relative entropy $D(\zeta||\tilde\rho_\text{f})$ of the initial thermal state with respect to the final state $\tilde\rho_\text{f}$ of this backward evolution. This is also intuitive: the lower our ability to time-reverse the evolution, the larger the entropy production.

Finally, the concept of EP can be extended at the level of single realizations, that corresponds to two-point quantum trajectories in the present quantum Jarzynski's protocol. Each trajectory $\gamma$ is defined by the outcomes of energy measurements performed at the beginning and at the end of the forward protocol, while $\tilde \gamma$ stands for its time-reversed counterpart, as introduced in the pioneering two-point energy measurement (TPEM) scheme~\citep{Esposito2009, Campisi2011, Binderbook2018}. The stochastic EP is defined as $\ep[\gamma] = \ln (p(\gamma)/p(\tilde \gamma))$ and compares the probability $p(\gamma)$ for $\gamma$ to be realized in the forward protocol and the probability $p(\tilde\gamma)$ for the corresponding $\tilde \gamma$ in the backward protocol \cite{Landi2020}. This expression provides us with another intuitive way to quantify irreversibility at the level of single trajectories. Although $\ep[\gamma]$ can take negative values, its average $\EP{}$ over all possible trajectories is non-negative by convexity of the exponential, in agreement with the SLT.

From this brief review, it appears that EP can be captured owing to various operational resources and, in particular, to the ability to access average or stochastic physical quantities as well as to run an evolution forward and backward. In what follows, we systematically employ these different approaches of EP to a basic protocol of an information-powered fridge. More specifically, the system is measured by a controller (readout $R$) and its further unitary evolution $U^{(k)}$ is set by the readout outcome $k$, thus leading to several different evolution branches of the feedback-controlled system, see Fig.~\ref{fig:entropy production}(b). Along with the two measurements forming the TPEM scheme the readout  outcome also contributes to the definition of the quantum trajectories. For the sake of clarity, we first detail the non-autonomous description of the protocol, where the feedback uses information encoded on a classical memory of the external controller. Then, we focus on the autonomous description, \ie for a fully closed system as reported in Ref.~\cite{Luis2020}.

\subsection{Average evolution in the non-autonomous description} \label{subsec:trajectories}

\begin{figure}
    \begin{centering}
    \includegraphics[width=0.99\columnwidth]{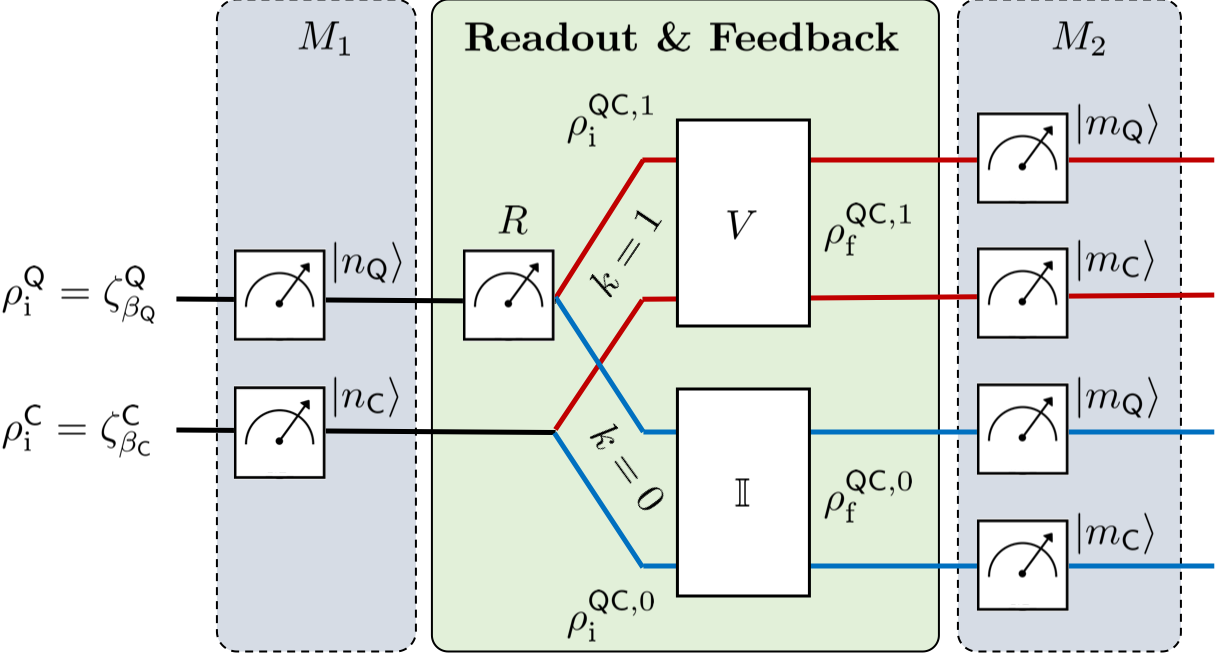}
    \par\end{centering}
    \caption{Non-autonomous system control with a binary readout completed with the two-point energy measurement (TPEM). The initial states of the systems \setup{Q} and \setup{C} are thermal states at different temperatures. The outcome $k=1$ of the demon readout $R$ sets a feedback interaction $V$ between \setup{Q} and \setup{C}. Otherwise, for $k=0$, \setup{Q} and \setup{C} do not interact, \ie they evolve under the identity $\mathbb{I}$. The TPEM is realized by two projective measurements, $M_1$ and $M_2$, in the energy basis of \setup{Q} and \setup{C} performed before and after the main protocol. Each trajectory is characterized by the set of five indices $\{\nQ,k,\nC,\mQ,\mC\}$, referring to the results of five measurements performed in a given protocol realization.}
    \label{fig:protocol}
    \end{figure}

Figure~\ref{fig:protocol} illustrates a non-autonomous description of the Maxwell's demon experiment studied in this paper. We consider a qubit \setup{Q} and a cavity \setup{C}. Their interaction is controlled by a third system further dubbed demon and denoted by \setup{D}. In this description, \setup{D} features a classical entity, performing a local projective measurement on \setup{Q} in its energy basis and storing its result in a classical memory. The two measurement outcomes are then exploited in the feedback loop (readout followed by feedback), that conditionally acts on the joint \setup{QC} system. Namely, they trigger a unitary system evolution $U^{(1)}=V$ for $k=1$ and no interaction, \ie the identity $U^{(0)}=\mathbb{I}$, for $k=0$. All the EP expressions derived in this and the next section are also valid for more general settings, with two arbitrary systems \setup{Q} and \setup{C}.

We first consider the average evolution of the joint \setup{QC} system and use the density matrix approach to describe its state. Initially, \setup{Q} and \setup{C} start at the local thermal equilibrium state $\rho_{\text{i}}^\fQC=\zeta_\betaQ^\fQ \otimes \zeta_\betaC^\fC$, where $\zeta_{\beta_{j}}^{j}=\exp[-\beta_{j}\left(H^{j}-F^{j}\right)]$ are the Gibbs states. For each system $j\in\{\fQ,\fC\}$, $H^{j}$ is the local Hamiltonian and $F^{j} = -(1/\beta_{j}) \ln\text{Tr}\big[e^{-\beta_{j}H^{j}}\big]$ is the equilibrium free energy. The internal energy of system $j$ in state $\rho^{j}$ is given by $\mathcal{U}^{j} = \text{Tr}_{j}\left[H^{j}\rho^{j}\right]$. Next, \setup{D} performs a projective measurement (\ie demon readout) on the system. The measurement outcome $k$ projects \setup{QC} onto the state $\rho_{\text{i}}^{\fQC,k}$ with probability $p(k)$. Then, \setup{D} stores the outcome $k$ and induces the unitary feedback operation $U^{(k)}$ between \setup{Q} and \setup{C} depending on $k$. There are thus several distinct branches, labelled by $k$, of the possible unitary evolution of the system. The final \setup{QC} states and their average over all measurement outcomes read  $\rhof^{\fQC,k} = U^{(k)} \rho_{\text{i}}^{\fQC,k}[U^{(k)}]^\dagger$ and $\rhof^{\fQC} = \sum_k p(k) \rhof^{\fQC,k}$, respectively. The relaxation of the non-equilibrium state  $\rhof^{\fQC}$ towards the initial thermal product state gives rise to the entropy production. The demon's memory, on the other hand, does not relax and hence does not produce entropy. This leads to our first expression of EP:
\begin{equation}
	\EP{1} = \Delta\beta {Q^\fC} +\mean{I},
	\label{eq::S1}
\end{equation}
where $\Delta\beta = \betaC-\betaQ$, $Q^\fC =\sum_{k}p(k)\Delta \mathcal{U}^{\fC,k}$ is the heat absorbed by \setup{C}, $\Delta \mathcal{U}^{\fC,k}$ is the energy change of \setup{C} during the feedback operation in branch $k$, and $\mean{I} = H[p(k)]=-\sum_k p(k)\ln p(k)$ is the Shannon entropy of the readout measurement. We use the fact that ${Q^\fQ} =-{Q^\fC}$ for a closed system and an energy-preserving readout, see Appendix~\ref{app:EP}. If there was no feedback action ($U^{(1)}=\mathbb{I}$), $\EP{1}$ would reduce to the well-known classical expression $\EP{} = \Delta\beta{Q}$ \citep{Landi2020}, quantifying the entropic counterpart of the heat exchanged between two systems. In addition to this exchange term, Eq.\eqref{eq::S1} explicitly involves an informational contribution. This is in agreement with the pioneering expressions of the SLT in the presence of a feedback control that were obtained by explicitly taking the demon's physical memory into account~\citep{Sagawa2008, Sagawa2009,Maruyama2009}. 

An alternative, second expression for the EP can be obtained starting from the following identity for an arbitrary state $\rho$: $D(\rho||\zeta) = \beta[\mathcal{U}(\rho)-F] -S(\rho)$. Writing the heat in \eqref{eq::S1} in terms of the quantum divergence we obtain
\begin{equation}
	\EP{2} = \sum_k p(k) \,D\!\left(\rhof^{\fQC,k} || \zeta_\betaQ^\fQ \otimes\zeta_\betaC^\fC \right),
	\label{eq::S2}
\end{equation}
where $\rhof^{\fQC,k}$ is the final $\fQC$ state conditioned on~$k$, see Appendix~\ref{app:EP}. This expression can be interpreted as follows. The entropy production for a thermalization process is known to be given by the quantum divergence of the initial state with respect to the final thermal one~\citep{Deffner2011}. For a given $k$, the entropy produced during the thermalization equals $D\!\left(\rhof^{\fQC,k} || \zeta_\betaQ^\fQ \otimes \zeta_\betaC^\fC\right)$. The total EP is, therefore, the average of such a conditional EP, associated to each branch $k$, over all readout outcomes.

The expressions $\EP{1}$ and $\EP{2}$ rely on the physical quantities provided by the forward protocol only. A third expression containing information also from the backward protocol can be obtained as well. For each branch $k$, the backward process is defined by the application of the time-reversed unitary $\tilde{U}^{(k)} = [U^{(k)}]^\dagger$ on the state after the thermalization, while the demon's memory remains unchanged. Thus, the probability of applying $\tilde{U}^{(k)}$ is given by the probability $p(k)$ of ending up in the forward branch $k$. Starting from \eqref{eq::S2} we show in Appendix~\ref{app:EP} that
\begin{equation}
	\EP{3} = \sum_k p(k)\, D\!\left(\rho_{\text{i}}^{\fQC,k} || \tilde \rho_\text{f}^{\fQC,k}\right),
	\label{eq::S3}
\end{equation}
where $\tilde\rho_\text{f}^{\fQC,k}$ is the \setup{QC} state of the backward protocol of the branch $k$ after the backward evolution $\tilde{U}^{(k)}$. This expression for the EP also comes in the form of the average over the outcomes of the readout measurement. It is a generalization of the equation obtained in Ref.~\citep{Batalhao2015}, where there is no feedback control being considered.

\subsection{Stochastic evolution in the non-autonomous description}\label{subsec:forward}

The system evolution can also be described stochastically by means of individual quantum trajectories. All thermodynamic quantities become trajectory-dependent, providing a finer description of the system dynamics. In the spirit of the TPEM scheme, the definition of our quantum trajectories involve the initial and final energy states of the joint \setup{QC} system, respectively denoted $\ket{\nQ,\nC}$ and $\ket{\mQ,\mC}$. In the present case where the dynamics generates no coherence in the energy basis, these states can be accessed by two energy measurements $M_1$ and $M_2$ respectively performed at the beginning and at the end of the feedback loop, of respective outcomes $\{\nQ,\nC\}$ and $\{\mQ,\mC\}$. 

The probability of the measurement outcomes $\{\nQ,\nC\}$ is $p(\nQ,\nC) = p(\nQ) p(\nC)$, where the probabilities $p(n_{j})$ are the Boltzmann weights of the initial uncorrelated thermal states. The demon readout R of outcome $k$ conditions the \setup{QC} evolution $U^{(k)}$ between $M_1$ and $M_2$. For the ideal readout, the state $\ket{\nQ}$ deterministically sets the value of $k$. In a more general case, we can consider a conditional probability $p(k |\nQ)$ of the readout outcome $k$ accounting for possible readout limitations, such as non-projective measurement or detection errors. Eventually, the trajectory $\gamma$ is defined by a unique set of the initial states $\ket{\nQ,\nC}$, the system evolution branch $k$, and the final state $\ket{\mQ,\mC}$. The forward trajectory probability distribution  $p(\gamma)$ is given by the probability of getting the set of outcomes $\gamma = \{\nQ,k,\nC,\mQ,\mC\}$ and it explicitly reads 
\begin{equation}
	p(\gamma) = p(\mQ,\!\mC|\nQ,\!k,\!\nC) \,p(k|\nQ) \,p(\nQ) \,p(\nC).
	\label{eq:trajectory probability forward}
\end{equation}

The total number of all possible trajectories is  $d_\fQ^2 \times d_\fC^2 \times d_\fQ$, where $d_j$ is the size of the Hilbert space of the system $j$. The last contribution, $d_\fQ$, comes from the fact that the external controller (demon) must contain $d_\fQ$ distinguishable states to encode the measurement outcomes.

    \begin{figure}
    \begin{centering}
    \includegraphics[width=0.99\columnwidth]{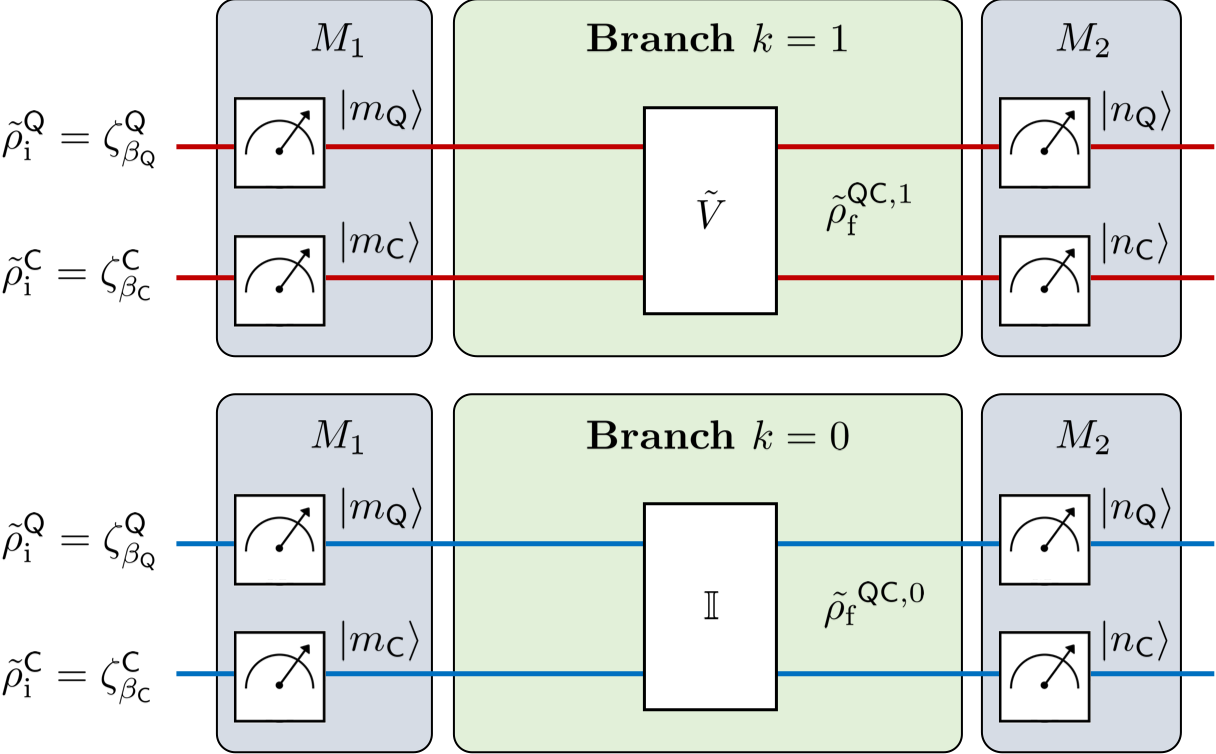}
    \par\end{centering}
    \caption{Backward trajectory protocol reversing the forward protocol of Fig.~\ref{fig:protocol} after thermalization. Two branches with $k=0$ and $1$ correspond to the two possible feedback evolutions in the forward protocol. 
    The backward unitary operation for $k=1$ is $\tilde U^{(1)} = \tilde V$ obtained by time-reversing $V$. For $k=0$, the identity operation is applied to the \setup{QC} system. The initial states in each branch are thermal states. Initial and final projective measurements, $M_1$ and $M_2$, are used for the TPEM. 
    \label{fig:backward trajectories}}
    \end{figure}

We now turn our attention to the distribution of the backward trajectories. As mentioned above, the backward process in each reverse branch $k$ is generated by the time-reversed unitary $\tilde{U}^{(k)}$. The backward trajectory $\tilde\gamma$ is defined by the set of parameters $\{\nQ, k, \nC, \mQ,\mC \}$ and is the counterpart of the forward trajectory $\gamma$ labelled by the same indices. Similarly to the forward protocol, the probability distribution of the backward trajectories is given~by 
\begin{equation}
	p(\tilde\gamma) = \pb(\nQ,\nC | \mQ,k, \mC) \,\pb(\mQ) \,\pb(\mC)\,p(k),
	\label{eq:trajectory probability backward}
\end{equation}
where $\pb(\nQ,\nC | \mQ,k, \mC)$ is the conditional probability of the final backward state. The initial backward probabilities $\pb(\mQ)$ and $\pb(\mC)$ are determined from the corresponding initial Gibbs states. The probability $p(k)$ of the backward branch $k$ equals the probability of the readout outcome $k$ in the forward protocol.

Given the probabilities of the forward trajectory $\gamma$ and of the corresponding backward trajectory $\tilde \gamma$, the stochastic EP is defined as $\sigma[\gamma] = \ln \big(p(\gamma)/p(\tilde\gamma)\big)$. The average EP computed over all $\gamma$'s is then given by~\citep{Landi2020}
\begin{equation}
	\EP{4}  = \sum_\gamma p(\gamma) \ln \frac{p(\gamma)}{p(\tilde \gamma) } = D\Big(p(\gamma) || p(\tilde{\gamma})\Big),
	\label{eq::S4}
\end{equation}
where the relative entropy $D$ is computed between probability distributions $p(\gamma)$ and $p(\tilde{\gamma})$. Note that for classical distributions $p_n$ and $q_n$, $D$ is defined by the Kullback-Leibler divergence: $D\left(p||q\right)= \sum_{n}p_{n} \ln\left(p_{n}/q_{n} \right)$~\citep{Coverbook2006}, which is equivalent to the quantum divergence between two states whose density matrices are diagonal in the same basis. The expression $\EP{4}$ quantifies the irreversibility by comparing the stochastic trajectories of the forward and backward protocols. Notably, its computation requires no knowledge on the actual physical states defining the trajectories, but needs only the ability to distinguish different trajectories in order to properly access their probabilities.

We define as $p\left(\sigma\right)=\sum_{\gamma}p\left(\gamma\right)\delta_{\sigma,\sigma\left[\gamma\right]}$ and $\pb(\sigma) =\sum_{\gamma} p\left(\tilde{\gamma}\right) \delta_{\sigma,\sigma\left[\gamma\right]}$ the total probability of the forward and backward trajectories, respectively, contributing to the value $\sigma$ of EP, where $\delta_{a,b}$ is the Kronecker delta. With these two probability distributions one can easily show the detailed fluctuation relation: $\exp(\sigma) =  p(\sigma) / \pb(\sigma)$. By averaging over all possible values of $\sigma$ we obtain
\begin{equation}
	\EP{5} = \sum_\sigma p(\sigma) \ln\frac{p(\sigma)}{\pb(\sigma)} = D\Big(p(\sigma) || \pb(\sigma)\Big).
	\label{eq::S5}
\end{equation}
The stochastic entropy production $\sigma[\gamma]$ for each trajectory $\gamma$, required for computing $\EP{5}$, can be obtained as
\begin{equation}
	\sigma[\gamma] = \betaQ Q^\fQ[\gamma]+\betaC Q^\fC[\gamma] + I[\gamma],
	\label{eq:stochastic}
\end{equation}
see Appendix~\ref{app:EP}. Here, $Q^{j}[\gamma]$ is the stochastic heat received by the system $j\in\{\fQ,\fC\}$ and $I[\gamma] = -\ln p(k)$ is the stochastic information extracted from the readout measurement. On the contrary to $\EP{4}$, the expression $\EP{5}$ compares the forward and backward probability contributions to the EP, even if $\sigma[\gamma]$ is degenerate for some trajectories. Thus, despite the obvious similarity of the mathematical expressions, $\EP{4}$ and $\EP{5}$ differ appreciably with respect to the information required for their computation.

\subsection{The autonomous description\label{subsec:average}}

\begin{figure}[t]
    \begin{centering}
    \includegraphics[width=0.78\columnwidth]{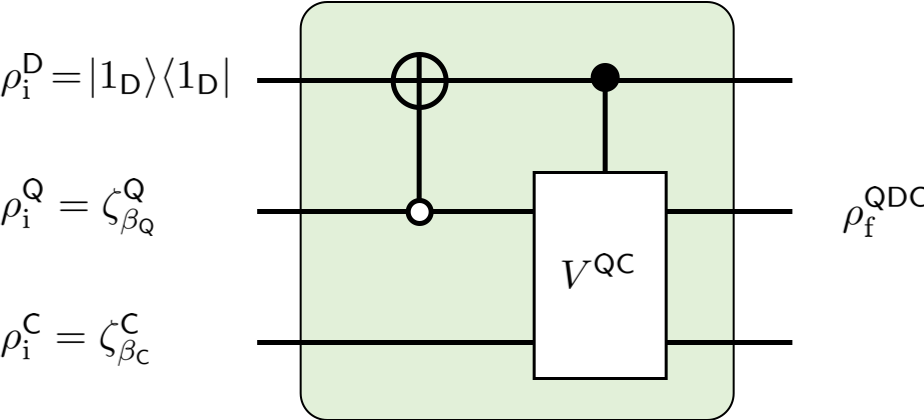}
    \par\end{centering}
    \caption{Quantum circuit of an autonomous Maxwell's demon. The demon \setup{D} and the system \setup{Q} are both qubits. The readout is represented by the controlled NOT gate with negative control line, \ie the state of \setup{D} is inverted if \setup{Q} is in state $\ket{0_\fQ}$. The controlled unitary implements the demon feedback. It switches on the \setup{QC} unitary interaction $V$ if the demon state is $\ket{1_\fD}\bra{1_\fD}$.
    \label{fig:circuit}}
    \end{figure}

In the non-autonomous description the demon \setup{D} has been treated as a classical feedback loop, involving a measurement and a conditional action on the \setup{QC} system. The demon's influence has been taken into account through the measurement outcome probability $p(k)$ quantifying the information extracted by \setup{D} from the system \setup{Q}. Alternatively, we can also consider a global \setup{QDC} system incorporating \setup{D} with the demon feedback action being part of a global unitary evolution of this closed system. For the remainder of this section we apply to our experiment the autonomous demon description reported in Ref.~\cite{Luis2020}. The equivalent quantum circuit for the forward protocol corresponding to a two-level (qubit) system \setup{Q} is depicted in Fig.~\ref{fig:circuit}. The demon, also assumed to be a qubit without loss of generality, starts in the pure reference state $\ket{1_\fD}$. Both the readout and the feedback operations are dynamically implemented by means of global unitaries on the total \setup{QDC} system. The projective readout in the energy basis of \setup{Q} is replaced by a controlled NOT (CNOT) gate. It transforms the initial state of \setup{D} into $\ket{0_\fD}$ if the state of \setup{Q} is $\ket{0_\fQ}$. After the CNOT gate, a controlled unitary operation between \setup{Q} and \setup{C} is performed to appropriately implement the feedback action. As expected, the reduced \setup{QC} state after this operation is $\sum_{k} p(k) \rhof^{\fQC,k}$, which is the average final state $\rhof^{\fQC}$ of the protocol described in Sec.~\ref{subsec:forward}.

Our final expression for the EP comes from the analysis of the closed \setup{QDC} system with the demon \setup{D} explicitly included as a third quantum system. We show in Appendix~\ref{app:EP} that 
\begin{equation}
	\Delta\beta Q^{\fC} = \Delta_{\text{fb}}I^{\fQC:\fD} + D\left(\rhof^\fQC || \zeta_\betaQ^\fQ \otimes \zeta_\betaC^\fC\right),
	\label{eq:equality correlations}
\end{equation}
where $I^{\fQC:\fD} = D\left(\rho^\fQDC || \rho^\fQC\!\otimes\! \rho^\fD\right)$ is the mutual information between \setup{QC} and \setup{D}, while $\Delta_\text{fb}$ denotes the information change during the feedback step, \ie before and after the controlled unitary gate. We consider here the ideal readout, \ie $p(k|\nQ)= \delta_{k,\nQ}$. This relation is a generalization to our current protocol of the expression first derived in Ref.~\citep{Luis2020}. It comes directly from the entropy conservation of the global \setup{QDC} system for the closed evolution depicted in Fig.~\ref{fig:circuit}. 

Since the correlations before the feedback step are given by the Shannon entropy $H[p(k)]$, substituting \eqref{eq:equality correlations} into \eqref{eq::S1} we obtain
\begin{equation}
	\EP{6} = D\left(\rhof^\fQC || \zeta_\betaQ^\fQ \otimes \zeta_\betaC^\fC\right) + I_{\text{f}}^{\fQC:\fD}.
		\label{eq::S6}
\end{equation}
It clearly shows that the EP has two contributions. The divergence quantifies the entropy produced in the thermalization process for the \setup{QC} system starting in the state $\rhof^\fQC$. The final mutual information between the two subsystems \setup{QC} and \setup{D} quantifies the amount of entropy produced by erasing all correlations between them, due to the thermalization. This result evidences that there is an entropic cost for erasing correlations~\cite{Landi2020,Camati2021}.

\subsection{Summary of all expressions}

Table~\ref{tab:EP} summarizes the six alternative expressions for the entropy production along with additional information on the underlying protocols and the statistical nature of the required physical quantities. Expressions $\Sigma_1$, $\Sigma_2$ and $\Sigma_6$ are based on the data extracted exclusively from the forward protocol. The other expressions require the execution and analysis of the backward protocol as well. We can distinguish three types of physical quantities showing up in different expressions: expressions $\Sigma_1$ and $\Sigma_6$ can be computed using information on the average initial and final states of the system (``averaged"). Expressions $\Sigma_2$ and $\Sigma_3$ are based on data averaged over different readout outcomes, thus requiring the discrimination of different evolution branches (``branched"). Finally, to compute expressions $\Sigma_4$ and $\Sigma_5$ we have to resolve individual trajectories (``stochastic").

Describing the same physical quantity, all these expressions are equivalent under the restriction of ideal unitary evolutions and ideal projective measurements, which have been used for their derivation. In the presence of realistic deviations from the idealized scenario, they start to differ, as will be shown in the next Section. For diagonal states, as considered here, only the expressions $\Sigma_2$ and $\Sigma_6$ stay mathematically identical and provide the same EP value irrespective of the evolution imperfections, see Appendix~\ref{app:EP26}.

\begin{table}
\centering
\resizebox{\columnwidth}{!}{
\begin{tabular}{lcc}
\hline \hline 
 \vphh 
 Expression & Protocol & Quantities  \\[5pt]
\hline 
$\Sigma_{1}=\Delta\beta Q^{\textsf{C}} +\left\langle I\right\rangle$ & forward & averaged \vph\\[10pt]
$\Sigma_{2}=\sum_{k}p\left(k\right)D\left(\rho_{\text{f}}^{\textsf{QC},k}||\zeta_{\beta_{\textsf{Q}}}^{\textsf{Q}}\otimes\zeta_{\beta_{\textsf{C}}}^{\textsf{C}}\right)$ & forward & branched \vph\\[10pt]
$\Sigma_{3}=\sum_{k}p\left(k\right)D\left(\rho_{\text{i}}^{\textsf{QC},k}||\tilde{\rho}_{\text{f}}^{\textsf{QC},k}\right)$ & \pbox{20cm}{forward,  \\ backward} & branched \vph\\[10pt]
$\Sigma_{4}=D\big(p\left(\gamma\right)||p\left(\tilde{\gamma}\right)\big)$ & \pbox{20cm}{forward,  \\ backward} & stochastic \vph\\[10pt]
$\Sigma_{5}=D\big(p\left(\sigma\right)||p_{\text{b}}\left(\sigma\right)\big)$ & \pbox{20cm}{forward,  \\ backward} & stochastic \vph\\[10pt]
$\Sigma_{6}=D\left(\rho_{\text{f}}^{\textsf{QC}}||\zeta_{\beta_{\textsf{Q}}}^{\textsf{Q}}\otimes\zeta_{\beta_{\textsf{C}}}^{\textsf{C}}\right)+I_{\text{f}}^{\textsf{QC:D}}$ & forward & averaged \vph\\[10pt]
\hline \hline 
\end{tabular}
}
\caption{The entropy production expressions.}
\label{tab:EP}
\end{table}

\section{Experimental results}

\subsection{Maxwell's demon system}

We measure the entropy production in the Maxwell's demon system described by the quantum circuit in Fig.~\ref{fig:circuit} and realized in a cavity QED setup~\cite{Luis2020}. Qubit \setup{Q} and two-level demon \setup{D} are simultaneously encoded into three adjacent circular Rydberg states of a single Rubidium atom \setup{A} (with principle quantum numbers $49$, $50$ and $51$ corresponding to atomic states $\ket f$, $\ket g$ and  $\ket e$, respectively). The mapping between the logical states of the \setup{QD} system and the physical states of \setup{A} is the following: $\ket{1_\fQ,\!1_\fD}=\ket{e}$, $\ket{0_\fQ,\!1_\fD} = \ket{g}$, and $\ket{0_\fQ,\!0_\fD} = \ket{f}$. According to the Maxwell's demon circuit, see Fig.~\ref{fig:circuit}, the state $\ket{0_\fQ,\!1_\fD}$ is never populated and does not need to be encoded in a particular physical state of \setup{A}. The system \setup{C} is realized with a high-quality superconducting microwave cavity resonant with the atomic $\ket{g}$--$\ket{e}$ transition at $51$~GHz and far detuned from the $\ket{g}$--$\ket{f}$ transition at $54$~GHz.

\begin{figure}[t]
     \includegraphics[width=\columnwidth]{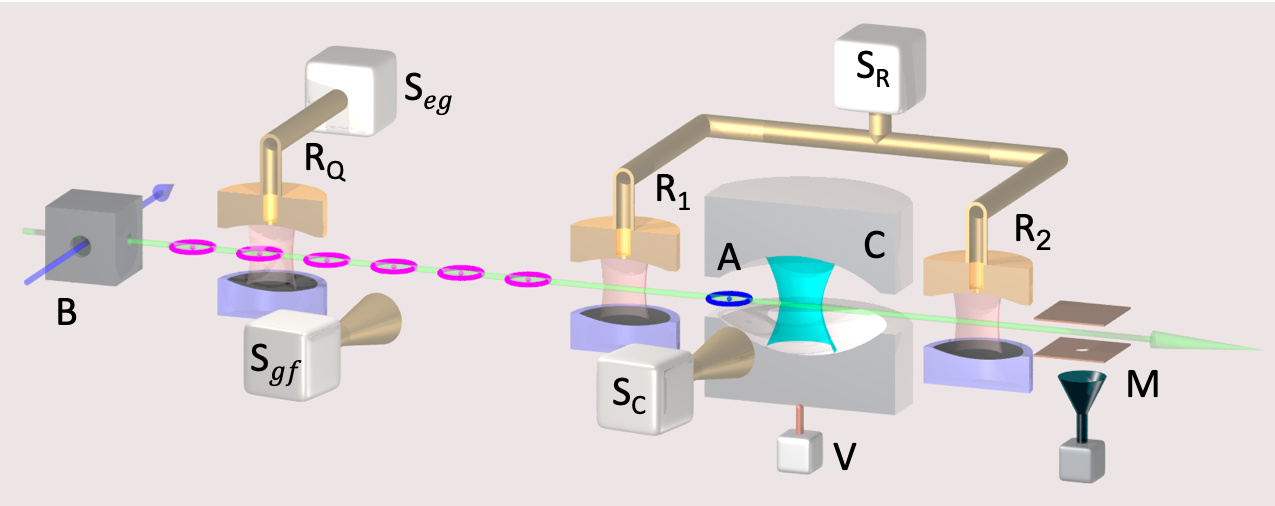}
     \caption{Schematic representation of the experimental setup. The Maxwell's demon system is realized with a microwave cavity (\setup{C}) and flying circular Rydberg atoms (blue toroid for the qubit-demon atom, magenta toroids for QND probe atoms). See text and Ref.~\cite{Luis2020} for details.}
     \label{fig:setup}
    \end{figure}

The basic experimental setup is schematically presented in Fig.~\ref{fig:setup}. Individual flying Rydberg atoms exit the preparation zone \setup{B} in the state $|g\rangle$. To prepare the state $|e\rangle$ a resonant microwave pulse is applied in \setup{R}$_\fQ$ by means of the microwave source \setup{S}$_{eg}$. Its amplitude and duration are adjusted to realize a Rabi $\pi$-pulse between $|g\rangle$ and $|e\rangle$. The demon readout is implemented by deterministically flipping the atomic states $\ket{g}$ and $\ket{f}$ before \setup{A} enters \setup{C}. This operation is induced by the microwave source \setup{S}$_{gf}$ resonant with the $\ket{g}$--$\ket{f}$ transition and adjusted to maximize the atomic population transfer.

The atom-cavity interaction is controlled by an electric field applied across \setup{C} by the voltage source \setup{V} via Stark-tuning the atomic frequency. The demon feedback is implemented by a resonant interaction between \setup{A}  and \setup{C} based on the adiabatic passage technique. It allows for the efficient population transfer between the \setup{AC} states $\ket{e,n}$ and $\ket{g,n+1}$ independent of the cavity photon number $n$. Energy conservation prevents the coupling of the joint ground state $\ket{g,0}$ to other states.

The atomic states are directly measured by a field-ionisation detector \setup{M} providing us with the final qubit and demon states, $\ket{\mQ}$ and $\ket{\mD=k}$, respectively. The cavity photon-number state $\ket{\mC}$ is probed by a sequence of several tens of atoms interacting with \setup{C} in the dispersive regime and performing a quantum non-demolition (QND) measurement of its photon number~\cite{Guerlin07}.

\subsection{Experimental sequences}
    
Each of the six EP expressions can be experimentally accessed by running the Maxwell's demon circuit and measuring physical quantities entering these expressions using measurement strategies properly adapted to each quantity. For instance, the cavity energy change $Q^\fC$, required to compute $\EP{1}$, can be obtained by comparing the average initial and final photon number in the cavity without the need to resolve different numbers. However, in order to significantly reduce the overall data acquisition time and to address all expressions at once we have decided to record the complete statistics of individual trajectories in the forward and backward protocols of the Maxwell's demon circuit. Knowing the initial and final states of each trajectory as well as their occurrence probability allows us to compute any physical quantity appearing in the EP expressions, as will be shown below.

In order to get the EP for any initial temperature of \setup{Q} and \setup{C} without increasing the overall experimental time, we have decided to replace the combination of the initial thermal state preparation and the first projective measurement of the TPEM scheme with the direct preparation of the \setup{QDC} system in the pure energy eigenstate $\ket{\nQ,1_\fD,\nC}$. The different thermal states are then taken into account by using the corresponding theoretical probability distributions $p(\nQ)$ and $p(\nC)$. We have shown in Ref.~\cite{Luis2020} that the experimental Gibbs' states of \setup{Q} and \setup{C} of given temperatures $\betaQ$ and $\betaC$ can be experimentally prepared and measured to be in good agreement with the theoretical distributions $p(\nQ)$ and $p(\nC)$.

Summarizing, in our basic experimental sequence we initially prepare the \setup{QDC} system in the pure energy eigenstate $\ket{\nQ,1_\fD,\nC}$ and measure the probability of its final state $\ket{\mQ,k,\mC}$ after the feedback evolution. In this way we obtain the conditional probability $p(\mQ,k,\mC | \nQ,\nC)$ of the trajectory $\gamma = \{\nQ,k,\nC, \mQ,\mC\}$. Note that the initial demon state is always $\ket{1_\fD}$ and thus does not enter into the trajectory definition. Finally, for any temperature $\betaQ$ and $\betaC$ with the corresponding $p(\nQ)$ and $p(\nC)$ we compute $p(\gamma)$ using Eq.~\eqref{eq:trajectory probability forward}.

In this work we consider, without loss of generality, the constant cavity temperature of $2.8$~K and the qubit temperature varying such that the relative inverse temperature $\delta\tilde\beta=1-\betaQ/\betaC \in [-6,6]$. Since the populations of the photon-number states larger than 3 are negligible for this temperature (the mean thermal photon number is $0.71$), we restrict $\nC$ to values from $0$ to $3$ only. 

The vacuum state $\ket{\nC=0}$ of \setup{C} is prepared by sending through its mode a beam of resonant atoms in state $\ket{g}$. They absorb all photons from \setup{C} thus cooling it into the vacuum. The state $\ket{\nC=1}$ with one photon is excited from the vacuum by using one atom in state $\ket{e}$ and forcing it to resonantly emit a photon into \setup{C}. The preparation of larger photon-number states are realized by the QND projection of a small coherent field~\cite{Guerlin07}. We first inject into \setup{C} a coherent field with about $3$ photons on average. Then, we perform the QND measurement, randomly resulting in different photon-number states. Finally, we post-select and sort all trajectories with the initial projected states $\ket{\nC=2}$ and $\ket{\nC=3}$.  

The final \setup{QDC} state is measured independently on each ensemble of quantum trajectories with the same initial state $\ket{\nQ,\nC}$. The final detection of \setup{A} gives us the conditional probability $p(\mQ,k | \nQ,\nC)$. The cavity photon-number probability is reconstructed on the ensemble of trajectories~\cite{Metillon19} with the same initial and final \setup{QD} state. In this way we obtain the conditional distribution $p(\mC | \nQ,\nC,\mQ,k)$ and compute $p(\mQ,k,\mC | \nQ,\nC) = p(\mC | \nQ,\nC,\mQ,k)\,p(\mQ,k | \nQ,\nC)$. The procedure of the state preparation and detection, along with all measured probabilities, is presented in detail in Appendix~\ref{app:data}. The probability of the trajectory $\gamma = \{\nQ,\nC,\mQ,k,\mC\}$ for each $\betaQ$ then equals $p(\gamma) = p(\mQ,k,\mC | \nQ,\nC) \,p(\nQ) \,p(\nC)$. A similar procedure is realized to obtain the probability distribution $p(\tilde \gamma)$ of the backward trajectories. The set of probabilities $\{p(\gamma)\}$ and $\{p(\tilde \gamma)\}$ are used in the following to compute all expressions for the entropy production, as explained below for each EP expression.

\subsection{Measurement of entropy production}

\begin{figure}[t]
     \includegraphics[width=\columnwidth]{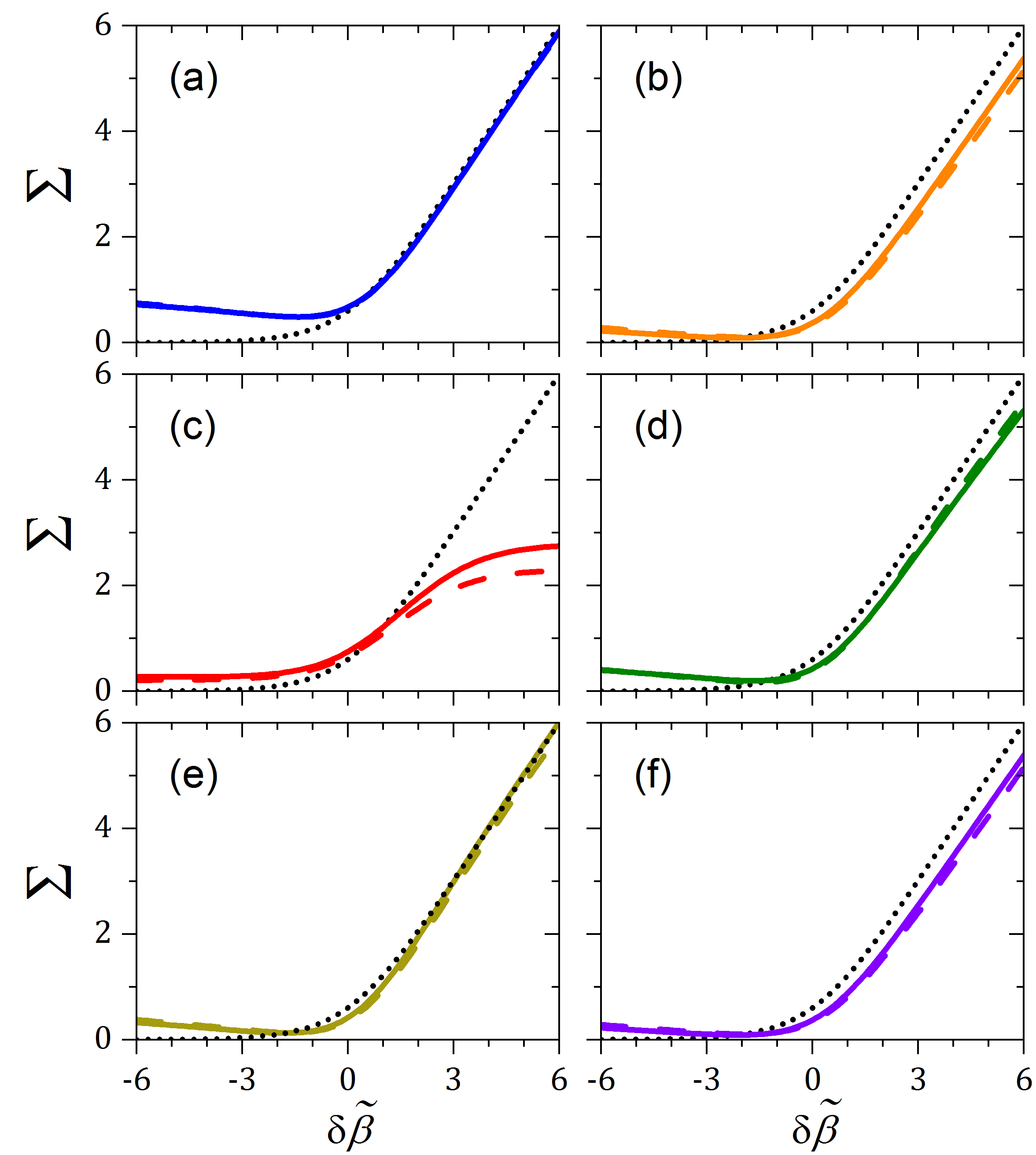}
     \caption{Entropy production calculated in natural units of information (nats) versus relative inverse temperature $\delta\tilde{\beta} = 1\!-\!\betaQ/\betaC$. Panels (a) to (f) correspond to the entropy production expressions  $\EP{1}$ to $\EP{6}$, respectively. Solid and dashed lines are computed from experimental and simulated data, respectively. Dotted lines are theoretical for the ideal model circuit of Fig.~\ref{fig:circuit} in the absence of any experimental imperfections.}
     \label{fig:results}
    \end{figure}

Figure~\ref{fig:results} shows the temperature dependence of the entropy production $\EP{}$ computed from the six expressions. Dotted lines correspond to the theoretical values for the ideal \setup{QDC} system as presented by the quantum circuit in Fig.~\ref{fig:circuit}. As expected, they coincide for all expressions, showing their fundamental equivalence. For large negative $\delta\tilde\beta$  (\ie the qubit state close to $\ket{0_\fQ}$), the probability for \setup{Q} to be in $\ket{1_\fQ}$ is small making the \setup{QC} interaction after the demon readout unlikely. In this limit, the \setup{QC} state stays almost unchanged reducing the entropy production to zero. For large positive $\delta\tilde\beta$ (\ie the qubit state close to $\ket{1_\fQ}$), $\EP{}$ linearly increases with $\delta\tilde\beta$, see Appendix~\ref{app:asymptotic}. Since \setup{Q} is mostly in $\ket{1_\fQ}$, the \setup{QC} interaction is extremely likely, pushing the \setup{QC} state further away from the initial thermal one and, consequently, producing more entropy.

The solid lines in Fig.~\ref{fig:results} are computed from the experimental results using the expressions $\EP{1}$ to $\EP{6}$ for panels (a) to (f), respectively. The deviation from the ideal curves is due to experimental imperfections. The most significant imperfections are the preparation error $\ePR$ of the initial atomic states, the errors of the readout ($\eRD$) and feedback ($\eFB$) operations and the discrimination error $\eDT$ of the atomic state measurement, see Appendix~\ref{app:major} for details. The errors $\eRD$ and $\eFB$ modify the system evolution. They change the entropy production and influence equally all experimentally obtained $\EP{}$. Namely, the imperfect readout allows for the non-negligible \setup{QC} interaction even for \setup{Q} prepared in $\ket{0_\fQ}$ resulting in the non-zero $\EP{}$ for $\delta\tilde\beta\ll0$. On the other hand, the imperfect feedback reduces the probability for the \setup{QC} interaction for \setup{Q} prepared in $\ket{1_\fQ}$, thus decreasing $\EP{}$ for $\delta\tilde\beta\gg0$. The errors $\ePR$ and $\eDT$ mix the labels of the detected quantum trajectories. Since different expressions are based on different combinations of experimental data, these errors have, in general, a different effect on the different expressions of $\EP{}$. Other imperfections, like atom and cavity relaxations, have a minor effect on the TPEM scheme and are listed in Appendix~\ref{app:minor}.

The computation of the first expression, $\EP{1}$, given by \eqref{eq::S1} and presented in Fig.~\ref{fig:results}(a), starts by computing the stochastic heat change $Q^\fQ[\gamma]$ of \setup{Q} for each trajectory $\gamma$. By averaging over all trajectories we get $Q^\fQ$. The probability $p(k=1)$ is given by the probability to finally detect \setup{D} in the state $\ket{1_\fD}$ and equals the sum of $p(\gamma)$ over all trajectories with $k=1$. Here, we have used only the data from the atomic state detection of the forward protocol (\ie no information on the cavity state is required). It is also noteworthy that the measured $\EP{1}$ is higher than $\EP{}$ based on other expressions. Ideally, the Shannon entropy $\mean{I}$ goes to zero for large negative and positive $\delta\tilde\beta$ when the demon state after the readout is a pure quantum state, $\ket{1_\fD}$ or  $\ket{0_\fD}$, respectively. However, due to the imperfect atomic state measurement $\eDT$, $\mean{I}$ is bound from below by $H[\eDT]$ thus shifting $\EP{1}$ up, as seen in Fig.~\ref{fig:results}(a).

The state $\rhof^{\fQC,k}$ in the expression $\EP{2}$ is obtained from the final probability distribution $p(\mQ,k,\mC)$. The product Gibbs state $\zeta_\betaQ^\fQ \otimes \zeta_\betaC^\fC $  is set by temperatures $\betaQ$ and $\betaC$. The probability $p(k)$ is obtained in the same way as for $\EP{1}$. Therefore, the current expression is based solely on the forward protocol after averaging the quantum trajectories into $\rhof^{\fQC,k}$. The experimental temperature dependence of $\EP{2}$ is shown in Fig.~\ref{fig:results}(b).

Figure~\ref{fig:results}(c) presents the expression $\EP{3}$ based on the analysis of the backward protocol with two branches, $k\!=\!0$ and $k\!=\!1$. It mainly relies on the backward trajectories, except for the value of $p(k)$ for the demon state extracted from the forward protocol. This expression shows the largest deviation from the ideal case for $\delta\tilde\beta\gg0$, which can be explained by the use of the backward trajectories and the divergent properties of $D$. The relative entropy $D(\rho||\sigma)$ is very sensitive to the smallest state variations if the support of the matrix $\sigma$ does not include the support of $\rho$, hence the second name ``divergence" for $D$. In the expression $\EP{2}$ the support of the reference state $\zeta_\betaQ^\fQ \otimes\zeta_\betaC^\fC$ is the whole Hilbert space of the \setup{QC} system, making this expression less sensitive to the small state variations. For the expression $\EP{3}$, however, the situation is radically different: both states appearing in the function $D$ have limited supports making its evaluation more sensitive to most experimental imperfections than all other expressions (see Appendix~\ref{app:major} for details).

The expression $\EP{4}$ in \eqref{eq::S4} is directly computed from the sets of $\{p(\gamma)\}$ and $\{p(\tilde\gamma)\}$ and is shown in Fig.~\ref{fig:results}(d).  It is the only expression based on all data measured in the forward and backward protocols with no additional transformation or averaging.

Figure~\ref{fig:results}(e) shows the relative entropy $\EP{5}$ obtained from \eqref{eq::S5}. We first compute, for each trajectory $\gamma$, the stochastic entropy production $\sigma[\gamma]$ from its initial and final state using \eqref{eq:stochastic}. Then, we calculate the probabilities $p(\sigma)$ and $\pb(\sigma)$ from the set of all values of $\sigma$ detected in the forward and backward protocols and obtain $\EP{5}$. This expression uses all experimental data after having grouped trajectories with the same $\sigma$. 

Finally, the expression $\EP{6}$ defined in \eqref{eq::S6} is shown in Fig.~\ref{fig:results}(f). The state $\rhof^\fQC$ is computed from the joint \setup{QDC} state $\rhof^\fQDC$, based on the distribution $p(\mQ,k,\mC)$, by tracing out \setup{D}. The mutual information between \setup{QC} and \setup{D} is computed directly on $\rhof^\fQDC$. Remarkably, the value of $\EP{6}$ perfectly coincides with that of $\EP{2}$. We show in Appendix~\ref{app:EP26} that these two expressions are mathematically identical and are based on the same set of the experimentally obtained physical quantities.

The dashed lines in Fig.~\ref{fig:results}  are the entropy productions computed from simulated data obtained by taking into account all mentioned experimental imperfections. The good agreement between the measurement and the simulation allows us to test and confirm the influence of various system's errors onto different ways to experimentally access the entropy production $\EP{}$. Some errors perturb quantum trajectories for particular temperature ranges. For instance, $\eRD$ manifests itself for $\delta\tilde\beta\ll 0$, while $\ePR$ and $\eFB$ are noticeable mainly for $\delta\tilde\beta\gg 0$. The detection error  $\eDT$ is influential for the qubit temperatures with very different populations in $\ket{0_\fQ}$ and $\ket{1_\fQ}$, \ie for $|\delta\tilde\beta| \gg 1$. The influence of other sources of errors on the discrepancy between the ideal and realistic cases depend on a particular expression for $\EP{}$ and on the way it is measured experimentally. In general, the errors increase $\EP{}$ for $\delta\tilde\beta \ll 0$ and decrease it for $\delta\tilde\beta \gg 0$ relative to the ideal case.

\section{Conclusion}

Our results allow to clarify the meaning of entropy production. Beyond its usual acception as a quantifier of irreversibility, it relates to some experimental lack of control over a quantum system, the larger the entropy production, the smaller the control.

In this spirit, we have presented different alternative ways to address and describe an ultimate information-powered quantum fridge, providing us different operational expressions for entropy production. Our cavity QED setup has allowed us to formulate theoretically and to access experimentally several expressions for $\EP{}$, each of them having its own physical interpretation. Their computation is based on different data and requires different data processing. However, describing the same physical quantity, they provide equivalent strategies to measure $\EP{}$. Following the same line, similar sets of entropy production expressions can be derived for any other system under investigation and characterization. The final experimentalist's choice is set by features and imperfections of a particular setup, perturbing measured data and thus the different $\EP{}$ expressions in different ways.

In the current work the state analysis has been restricted to the populations of energy states, sufficient for accessing the entropy production of the thermalization. To study other types of environments, it might be necessary to access quantum information \eg stored in the system's coherence or entanglement between its parts. Our experimental setup allows for the complete quantum state tomography \cite{Metillon19} providing access to quantum information and its transformation. We plan to use this ability and to implement dephasing and decorrelating environments in the forward-reservoir-backward protocol in order to reveal how different types of the information erasure induce irreversibility.

\begin{acknowledgments}

We thank J.-M. Raimond and M. Brune for insightful discussions.
We acknowledge support by European Community (SIQS project) and by the Agence Nationale de la Recherche (QuDICE project). P.~A.~C. acknowledges Templeton World Charity Foundation, Inc. This publication was made possible through the support of the grant TWCF0338 from Templeton World Charity Foundation, Inc. The opinions expressed in this publication are those of the author(s) and do not necessarily reflect the views of Templeton World Charity Foundation, Inc.
\end{acknowledgments}

\appendix

\section{Derivation of the different expressions of entropy production}\label{app:EP}

\subsection*{Expressions $\EP{1}$ and  $\EP{5}$}

The derivation of the expressions $\EP{1}$ and $\EP{5}$ are based on the definition of the stochastic entropy production given by Eq.~\eqref{eq:stochastic} of the main paper. Here we derive this equation from Eqs.~\eqref{eq:trajectory probability forward} and \eqref{eq:trajectory probability backward}. Since the systems $\fQ$ and $\fC$ start in the thermal state, the initial probabilities for the energy measurement are given by $p(\nQ)=\exp\{-\betaQ (E_{\nQ }^\fQ-F^\fQ )\} $ and $p(\nC)=\exp\{ -\betaC(E_{\nC}^\fC-F^\fC)\} $, where $E_\nQ^\fQ$ and $E_\nC^\fC$ are the energy eigenvalues of $H^\fQ $ and $H^\fC$, respectively. The readout is performed in the energy basis of $\fQ$. We assume at this moment that there is no readout errors, \ie $p(k|\nQ)=\delta_{k,\nQ}$. The remaining conditional probability in Eq.~\eqref{eq:trajectory probability forward} reads
\begin{equation}
	p(\mQ,\!\mC | \nQ,\!k,\!\nC) = \left|\!\left\langle E_\mQ^\fQ,\!E_\mC^\fC\right | U^{\fQC,(k)}\!\left|E_\nQ ^\fQ,\!E_\nC^\fC\right\rangle \!\right|^2\!\!\!,
\end{equation}
where $\left|E_{\nQ }^\fQ\right\rangle $ and $\left|E_{\nC}^\fC\right\rangle $
are the energy eigenstates of $\fQ$ and $\fC$, respectively. If $k\neq \nQ$, the conditional probability $p\left(k|\nQ \right)$ makes the whole trajectory probability $p\left(\gamma\right)$ equal to zero and the corresponding stochastic entropy production will not contribute to the average. Hence, it makes sense only to compute $p\left(\gamma\right)$ for which $k=\nQ$. Note that we still keep the two indices separately, as it is done in the main text.

For the backward trajectory probability $p(\tilde\gamma)$ in Eq.~\eqref{eq:trajectory probability backward}, the branch probability is given by $p(k)=p(\nQ)$, while the initial probabilities are $\pb(\mQ ) = \exp\{ -\betaQ (E_\mQ^\fQ -F^\fQ )\} $ and $\pb(\mC) = \exp\{ -\betaC (E_\mC^\fC-F^\fC)\} $. The remaining conditional probability in Eq.~\eqref{eq:trajectory probability backward} is given by 
\begin{equation}
	\pb(\nQ,\!\nC | \mQ,\!k,\!\mC) = \left|\!\left\langle E_\nQ^\fQ, \!E_\nC^\fC \right|\tilde{U}^{\fQC,(k)}\left |E_\mQ^\fQ,\!E_\mC^\fC \right\rangle \!\right|^{2}\!\!\!.
\end{equation}
By computing the stochastic entropy production from the ratio of $p(\gamma)$ to $p(\tilde\gamma)$ and considering the ideal measurement case, we obtain 
\begin{equation}
	\Sigma[\gamma] = \betaQ \left(E_\mQ^\fQ-E_\nQ^\fQ\right)+\betaC \left(E_\mC^\fC-E_\nC^\fC\right)-\ln p(k).\label{eq:stochastic entropy appendix}
\end{equation}
Since there is no source of work in this dynamics we identify $Q^\fQ [\gamma]=E_\mQ^\fQ-E_\nQ^\fQ$ and $Q^\fC[\gamma]=E_\mC^\fC-E_\nC^\fC$ as the stochastic heat absorbed by $\fQ$ and $\fC$, respectively. Defining $I[\gamma] = -\ln p(k)$, we arrive at Eq.~\eqref{eq:stochastic}.

\subsection*{Expression $\EP{2}$}

We present the derivation of the expression $\EP{2}$, given in Eq.~\eqref{eq::S2}, starting from Eq.~\eqref{eq::S1}. The total heat absorbed by $\fQ$ and $\fC$ can be rewritten in terms of the states $\rho_{\text{i}}^{\fQC,k}$ and $\rhof^{\fQC,k}$ as $Q^\fQ =\sum_{k}p(k)\Delta \mathcal{U}^{\fQ,k}$ and $Q^\fC =\sum_{k}p(k)\Delta \mathcal{U}^{\fC,k}$, respectively. Here $\Delta \mathcal{U}^{\fQ,k}=\text{Tr}_\fQ \left[H^\fQ \left(\rhof^{\fQ,k}-\zeta_{\betaQ }^\fQ \right)\right]$ is the energy change of $\fQ$ for the branch $k$, and similarly for $\Delta \mathcal{U}^{\fC,k}$. Now, writing these energy changes for each branch and employing the divergence property we obtain
\begin{eqnarray}
	\beta_j \Delta \mathcal{U}^{j,k}&=&\Delta S^{j,k}+D\left(\rhof^{j,k}||\zeta_{\beta_j}^j \right)
	\label{eq:4}
\end{eqnarray}
with 
$\Delta S^{j,k}=S\left(\rhof^{j,k}\right)-S\left(\zeta_{\beta_j}^j \right)$ and $j\in\{\fQ,\fC\}$. Next, we add Eq.~(\ref{eq:4}) for \setup{Q} and \setup{C}, and use the following two identities:
\begin{equation}
\Delta S^{\fQ,k}+\Delta S^{\fC,k}=S\left(\rhof^{\fQC,k}\right)+I\left(\rhof^{\fQC,k}\right)-S(\zeta),
\end{equation}
where we denote $\zeta = \zeta_\betaQ^\fQ \otimes\zeta_\betaC^\fC$ for simplicity, and 
\begin{equation}
	D\!\left(\rhof^{\fQC,k}||\zeta\right) \!=\! D\!\left(\rhof^{\fQ,k}||\zeta_{\betaQ }^\fQ \right)\!+\!D\!\left(\rhof^{\fC,k}||\zeta_{\betaC}^\fC\right)\!+\!I\!\left(\rhof^{\fQC,k}\right).
\end{equation}
We get 
\begin{equation}
	\betaQ \Delta \mathcal{U}^{\fQ,k}+\betaC\Delta \mathcal{U}^{\fC,k} = S\!\left(\rhof^{\fQC,k}\right) - S(\zeta)+D\!\left(\rhof^{\fQC,k}||\zeta\right).
\end{equation}
Since $S(\rhof^{\fQC,k}) = S(\rho_{\text{i}}^{\fQC,k})$ due to the unitary feedback operations and since $S(\rho_{\text{i}}^{\fQC,k})=S(\zeta_{\betaC}^\fC)$
because the feedback measurement is projective, we end up with 
\begin{equation}
	\betaQ \Delta \mathcal{U}^{\fQ,k}\!+\!\betaC\Delta \mathcal{U}^{\fC,k}\!+\!S\!\left(\zeta_{\betaQ }^\fQ \right) = D\!\left(\rhof^{\fQC,k}||\zeta\right).
	\label{eq:s4}
\end{equation}
For an ideal readout, $S(\zeta_\betaQ^\fQ)=\mean{I}$. Averaging \eqref{eq:s4} over $p(k)$ we finally obtain the equivalence between Eqs.~\eqref{eq::S1} and \eqref{eq::S2}.

\subsection*{Expression $\EP{3}$}

We derive the expression $\EP{3}$ of Eq.~\eqref{eq::S3} starting from Eq.~\eqref{eq::S2}. Since the divergence is invariant under unitary transformations, \ie $D(U\rho U^{\dagger}||U\sigma U^{\dagger})=D(\rho||\sigma)$, using the definition of $\rhof^{\fQC,k}$ we obtain 
\begin{equation}
\begin{split}
	D\left(U^{\fQC,(k)}\rho_{\text{i}}^{\fQC,k}\left[U^{\fQC,(k)}\right]^{\dagger}||\zeta\right) = \\
	D\left(\rho_{\text{i}}^{\fQC,k} || \left[U^{\fQC,(k)}\right]^\dagger \zeta U^{\fQC,(k)}\right).
\end{split}
\label{eq:6}
\end{equation}
The very last state is the definition of $\tilde \rho_\text{f}^{\fQC,k}$, \ie the state of the $k$th branch of the backward protocol after performing the unitary $\tilde{U}^{\fQC,(k)}$. By averaging Eq.~(\ref{eq:6}) we get Eq.~\eqref{eq::S3}.

\subsection*{Expression $\EP{6}$}

We start to derive the expression $\EP{6}$ given in Eq.~\eqref{eq::S6} by directly substitutng Eq.~\eqref{eq:equality correlations} into \eqref{eq::S1} and obtain
\begin{equation}
 \EP{6} =\Delta_{\text{fb}}I^{\textsf{QC:D}}+D\left(\rhof^{\fQC}||\zeta\right)+\mean{I}.
\end{equation}
Since $\mean{I}$ is the mutual information just before the implementation of the feedback~\cite{Luis2020}, we obtain $\Delta_{\text{fb}}I^{\textsf{QC:D}}+\mean{I} = I_{\text{f}}^{\textsf{QC:D}}$.

\section{Mathematical equivalence of the expressions $\EP{2}$ and $\EP{6}$ for diagonal states}\label{app:EP26}

The entropy production expressions $\EP{2}$ and  $\EP{6}$ are based on the same data treatment and are thus mathematically equivalent. In order to show this, we first remind several basic definitions from the information theory. The conditional entropy is defined as
\begin{eqnarray}
	H(A|B) &=& H(AB)-H(B),\label{eq:HAB1}\\
	H(A|B) &=& \sum_{a,b} p(a,b) \ln \frac{p(b)}{p(a,b)},\label{eq:HAB2}
\end{eqnarray}
where $A$ and $B$ are two random variables with the probability distributions $p(a)$ and $p(b)$, respectively. The probability $p(a,b)$ is the joint one defined through the conditional probability $p(a|b)$ as
\begin{equation}\label{eq:pxy}
	p(a,b) = p(a|b)\, p(b).
\end{equation}
Finally, for diagonal density operators used in the current paper, the complete density matrix $\rhoQC$ is computed from the branched ones, $\rhoQCk$, as
\begin{equation}\label{eq:rhoQC}
	\rhoQC = \sum_k p(k) \rhoQCk.
\end{equation}

We start by computing the difference of the two expressions:
\begin{widetext}
\begin{eqnarray}\label{eq:Y}
\EP{2}-\EP{6}
&=& \sum_{k} p(k) \Big\{ -H(\rhoQCk) - \Tr[\rhoQCk \ln\zeta] \Big\}  + H(\rhoQC) + \Tr [\rhoQC\ln\zeta] - \I 	\quad\quad\textrm{(from the definition of $D$)}\nonumber\\
&=& -\left\langle H(\rhoQCk)\right\rangle - \Tr \Big[ \sum_{k} p(k) \rhoQCk \ln \zeta \Big] 
 +H(\rhoQC) + \Tr\Big[ \sum_{k} p(k) \rhoQCk \ln\zeta \Big] - \I  \quad\quad\textrm{[from \eqref{eq:rhoQC}]}\nonumber\\
&=& -\left\langle H(\rhoQCk)\right\rangle + H(\rhoQC) - \Big(H(\rhoQC)  +H(\rhoD) - H(\rhoQCD)\Big)\nonumber\\
&=& -\left\langle H(\rhoQCk)\right\rangle + H(\rhoQCD)  - H(\rhoD)
= H(\rhoQC|\rhoD) - \left\langle H(\rhoQCk)\right\rangle  \quad\quad\textrm{[from \eqref{eq:HAB1}]}.
\end{eqnarray}
\end{widetext}

Next, we develop $\langle H(\rhoQCk)\rangle$:
\begin{widetext}
\begin{eqnarray}\label{eq:X}
	\langle H(\rhoQCk)\rangle 
	&=& \sum_k p(k) H(\rhoQCk) 
	= - \sum_k p(k) \Tr [\rhoQCk \ln\rhoQCk]\nonumber\\
	&=& - \sum_k p(k) \sum_{\fQC} p(\fQC|k) \ln p(\fQC|k) \quad\quad\textrm{[use the diagonal form of } \rhoQCk]\nonumber\\
	&=& - \sum_{\fQC ,k} p(k) p(\fQC|k) \ln p(\fQC|k)	\quad\quad\textrm{[from \eqref{eq:pxy}]}\nonumber\\
	&=& \sum_{\fQC ,k} p(\fQC k) \ln \frac{p(k)}{p(\fQC k)} 
	= H(\rhoQC|\rhoD)  \quad\quad\textrm{[from \eqref{eq:HAB2}]}.
\end{eqnarray}
\end{widetext}
By inserting \eqref{eq:X} into \eqref{eq:Y}, we finally get $\EP{2}=\EP{6}$. Note that the only constraint used here is the diagonal form of all density operators, \ie possessing no quantum coherences. Under this condition the von Neumann and Shannon entropies are equal merging together the classical probability description and the quantum density matrix formalism. No other assumptions on the origin of $\rhof^{\fQDC}$ and its reduced states, as well as no information on the demon action have been used.

\section{Asymptotic behaviour of entropy production}\label{app:asymptotic}

The simplest way to obtain the asymptotic behaviour of the entropy production $\EP{}$ in the limit of large relative inverse temperature $\delta\tilde\beta$ is to consider the expression $\EP{1}$, which can be rewritten as 
\begin{equation}
	\EP{1} = \Big( \betaC Q^\fC \Big)\, \delta\tilde\beta +\mean{I}.
	\label{eq::S1app}
\end{equation}

When $\delta\tilde\beta\rightarrow+\infty$, the qubit state approaches $\ket{1_\fQ}$ and therefore the mean heat transfer $Q^\fC \rightarrow \hbar \omegaC$, where $\hbar\omegaC$ is the energy of one cavity photon. At the same time the demon state at the end of the protocol approaches the pure state $\ket{0_\fD}$ forcing $\mean{I}\rightarrow 0$. Consequently, 
\begin{equation}
	\EP{1} \xrightarrow{\delta\tilde\beta\rightarrow+\infty}  \Big(\betaC \hbar\omegaC\Big) \, \delta\tilde\beta.
	\label{eq::S1limit}
\end{equation}

When $\delta\tilde\beta\rightarrow-\infty$, both the heat transfer $Q^\fC$ and the Shannon entropy $\mean{I}$ tend to zero. Thus, the entropy production $\EP{}$ does not depend on the temperature any more and converges to zero:
\begin{equation}
	\EP{} \xrightarrow{\delta\tilde\beta\rightarrow-\infty}  0.
	\label{eq::S1limit2}
\end{equation}

\section{Experimental data}\label{app:data}

The main experimental data used for this work are the conditional probabilities $p(\mQ,\!k,\!\mC|\nQ,\!\nC)$ measured for the forward trajectory protocol and $ \pb(\nQ,\nC | \mQ,k, \mC)$ for the backward trajectory protocol. All this data is given in an auxiliary ZIP archive \textit{TrajectoryProbabilities.zip} containing two ASCII files: \textit{Forward.dat} and \textit{Backward.dat}.

Rows and columns of the table \textit{Forward.dat} containing $p(\mQ,\!k,\!\mC|\nQ,\!\nC)$ correspond to the initial ($\ket{\nQ,\nC}$) and final ($\ket{\mQ,\!k,\!\mC}$) states of the \setup{QDC} system, respectively. The first column and the first row contain the labels of the corresponding states in the form `$(\nQ,\nC)$' and `$(\mQ,\!k,\!\mC)$', respectively. 

Rows and columns of the table \textit{Backward.dat} containing $ \pb(\nQ,\nC | \mQ,k, \mC)$ correspond to the final ($\ket{\nQ,\nC}$) and initial ($\ket{\mQ,\!k,\!\mC}$) states in the backward protocol, respectively. Similarly to the forward data table, the first column and row contain the labels of the corresponding states.

\section{Major experimental imperfections}\label{app:major}

The main experimental imperfections of our experimental setup affecting the measured entropy production are the following: 
\begin{itemize}
\item  the nonideal purity of the initial atomic state prepared for the TPEM protocols (error $\ePR$),
\item  the limited readout efficiency (error $\eRD$),
\item  the limited feedback efficiency (error $\eFB$),
\item  the imprecision of the final state measurement (error $\eDT$).
\end{itemize}
The errors $\eRD$ and  $\eFB$ modify the \setup{QDC} system evolution and, thus, the entropy production. Therefore, they influence $\EP{}$ computed from all expressions in the same way. On the other hand, the errors $\ePR$ and  $\eDT$ limit our ability to resolve different quantum trajectories of the system. They can have different effects to different expressions depending on how exactly they are computed. Below we comment on each of these errors. In the next Section we present other minor sources of experimental errors.

\afterpage{
\begin{figure}[h]
 	\includegraphics[width=0.91\columnwidth]{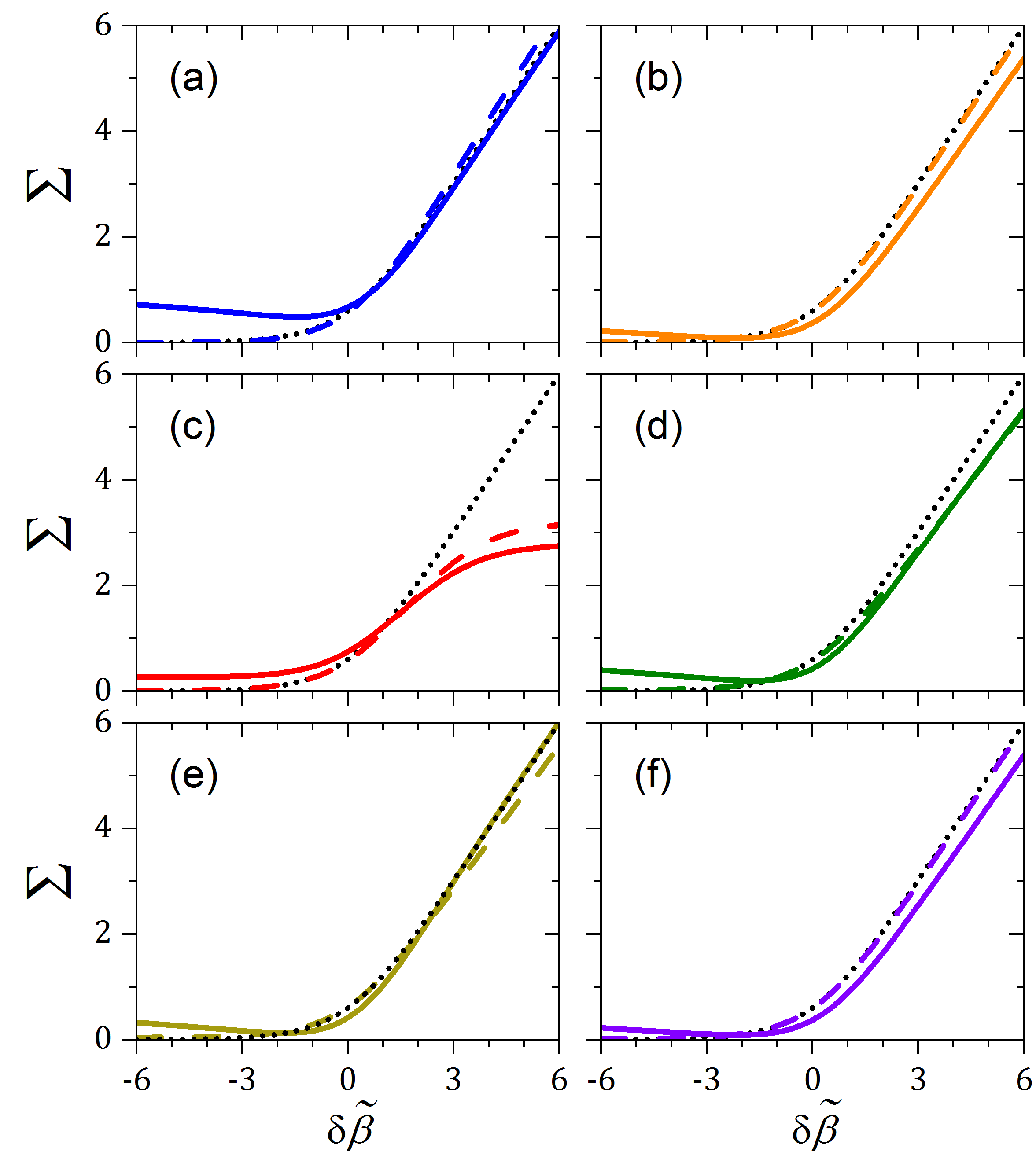}
	\caption{Simulation with only preparation error $\ePR$. 
}
	\label{fig:resultsPrep}
\end{figure}

\begin{figure}[h] 		
	\includegraphics[width=0.91\columnwidth]{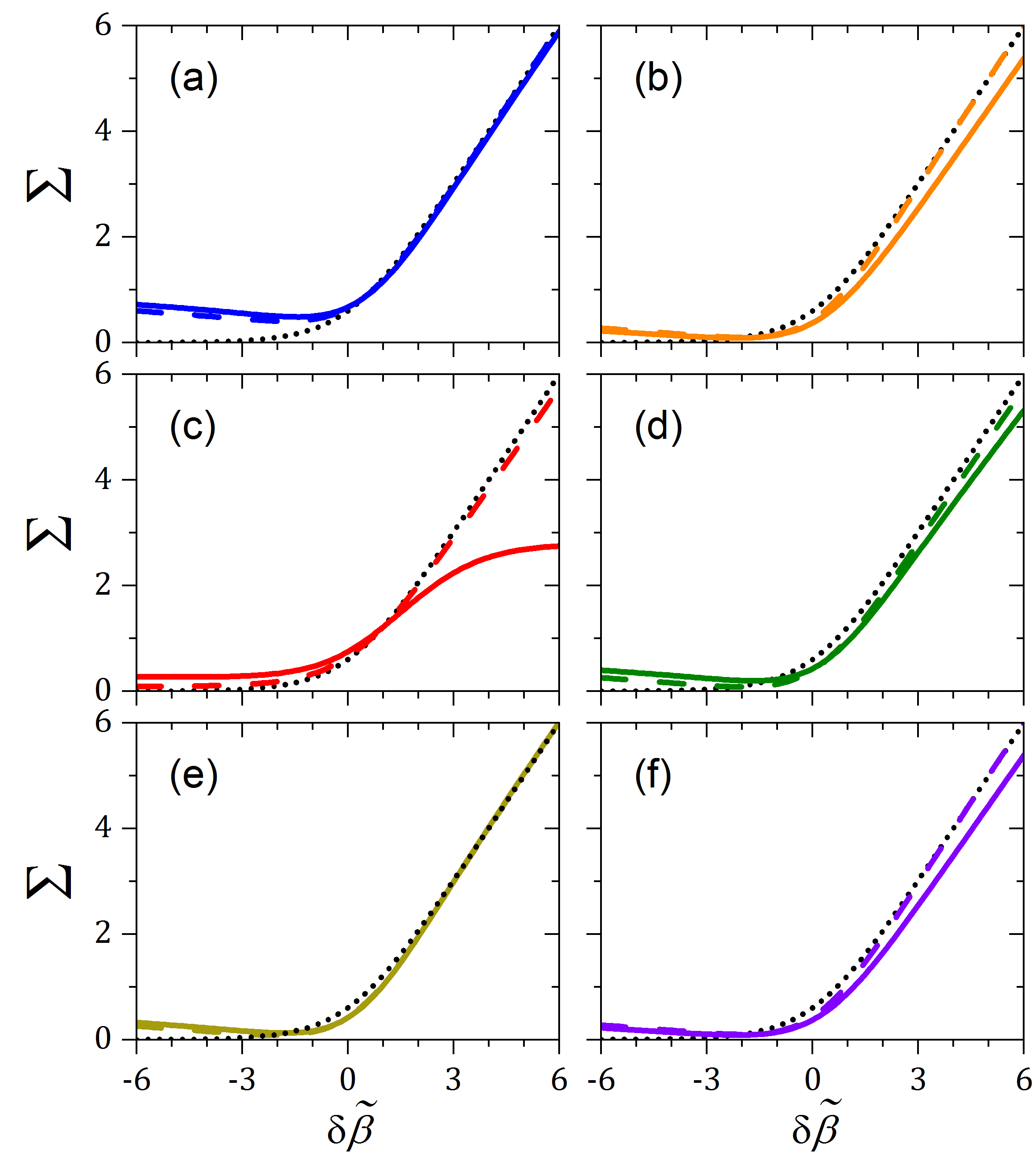}
 	\caption{Simulation with only readout error $\eRD$.}
 	\label{fig:resultsRead}
\end{figure}
	
\begin{figure}[h] 		
	\includegraphics[width=0.91\columnwidth]{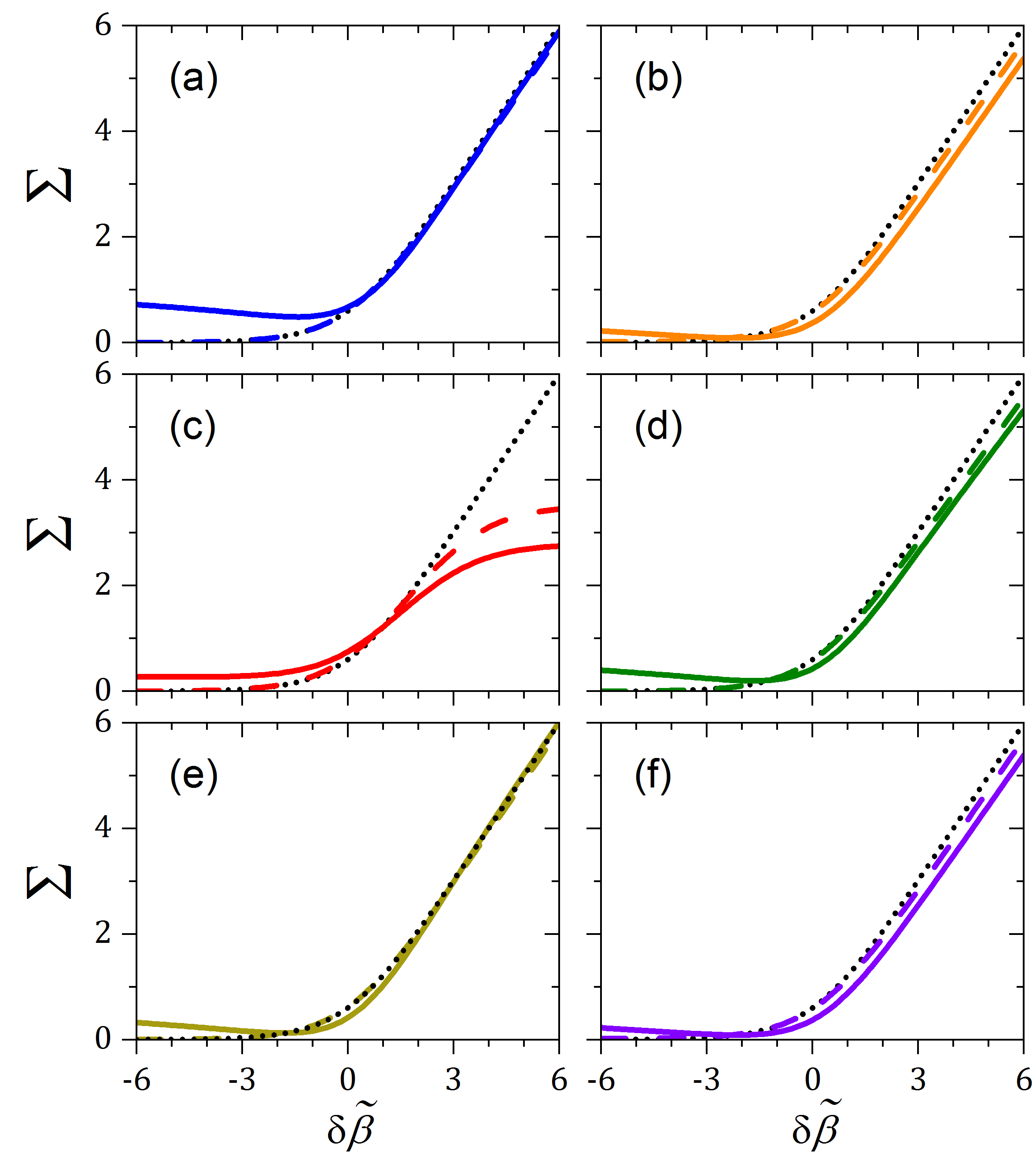}
 	\caption{Simulation with only feedback error $\eFB$.}
 	\label{fig:resultsFeed}
\end{figure}

\begin{figure}[h] 		
	\includegraphics[width=0.91\columnwidth]{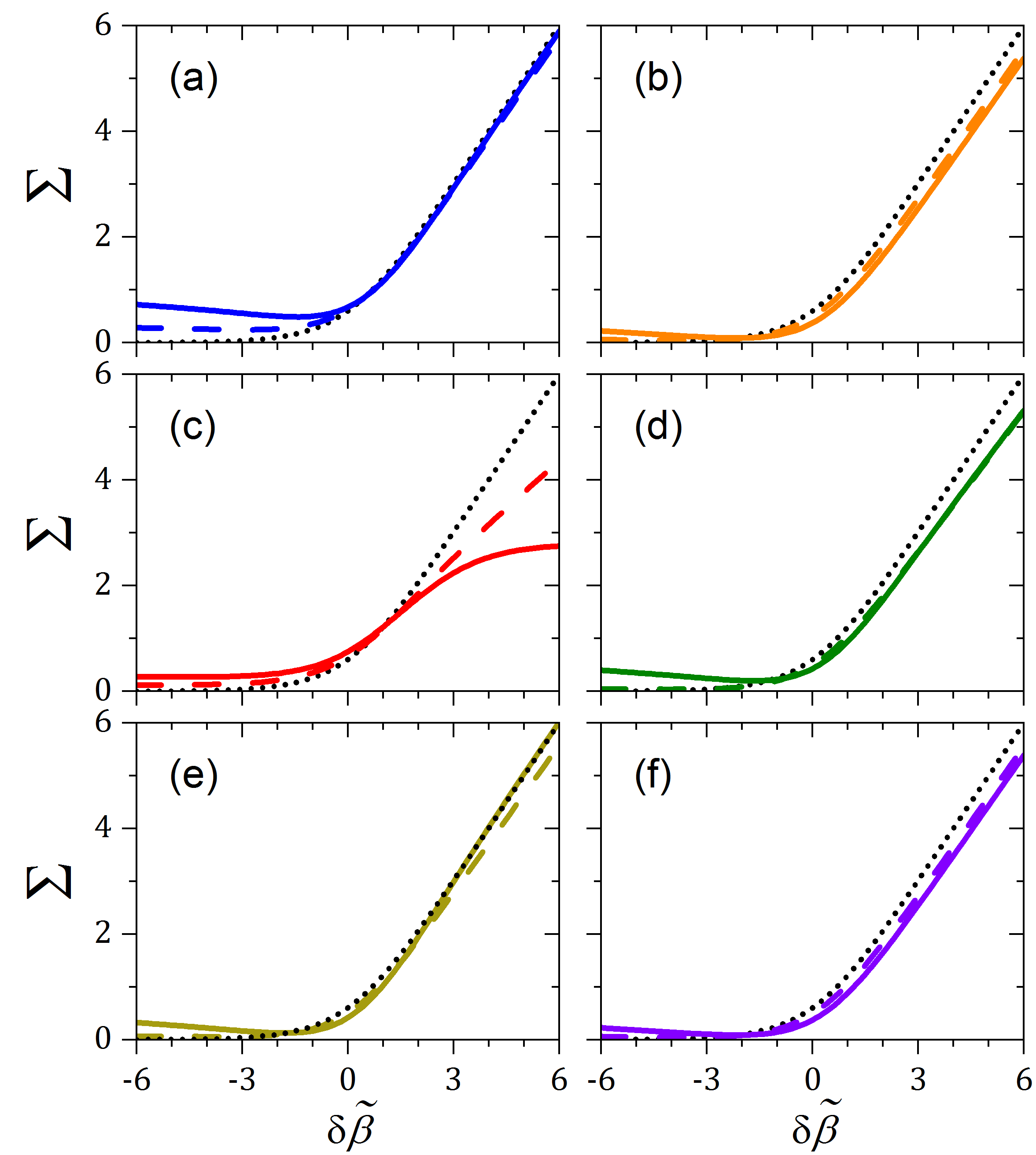}
 	\caption{Simulation with only state measurement error $\eDT$.}
 	\label{fig:resultsMeas}
\end{figure}
}

The preparation of the initial atomic state $\ket{g}$ is exact. The excited state $\ket{e}$, however, is prepared with the population $1-\ePR=0.9$, where $\ePR=0.1$ is the probability for the atom to be left in its ground state $\ket{g}$ due to the imperfect excitation pulse in \setup{R}$_\text{Q}$. Therefore, this error source is relevant for the temperature range $\delta\tilde\beta\gg0$ and is negligible for small qubit temperatures.

The demon readout is realized by a microwave $\pi$-pulse resonant with the atomic transition between levels $\ket{g}$ and $\ket{f}$. Being imperfect, this operation leaves the atom in its initial state $\ket{g}$ with the undesirable probability $\eRD=0.11$. Obviously, the readout does not affect the atom in the state $\ket{e}$. Therefore, this error source is relevant for low qubit temperatures ($\delta\tilde\beta\ll0$) and is negligible otherwise.

The feedback operation is a resonant population transfer between the atom and the cavity realized by means of the adiabatic passage technique. The population transfer has a failure probability of $\eFB=0.03$. Contrary to $\eRD$, the error $\eFB$ has effect on the system for $\delta\tilde\beta \gg0$ and is negligible otherwise.

The error of the final state measurement originates from the limited state resolution of our field-ionization detector in combination with the atomic relaxation during the atom flight from the cavity to the detector. The probabilities $\eDTx{a,b}$ to erroneously detect a state $\ket{b}$ as the state $\ket{a}$, with $a,b\in\{e,g,f\}$, have been independently measured to be: $\eDTx{e,f}=0.01$, $\eDTx{f,e}=0$, $\eDTx{e,g}=0.05$, $\eDTx{g,e}=0.02$, $\eDTx{g,f}=0.05$, and $\eDTx{f,g}=0.02$. The influence of these errors is different for different temperatures and different $\EP{}$ expressions.

To illustrate the effect of the major error sources onto the measured values of $\EP{}$, we have performed a series of numerical simulations with only one of the error sources activated. Figures~\ref{fig:resultsPrep} to \ref{fig:resultsMeas} present four simulations with the individual errors $\ePR$, $\eRD$, $\eFB$ and $\eDT$, respectively. The panels from (a) to (f) correspond to the expressions from $\EP{1}$ to $\EP{6}$. The line types are the same as in Fig.~\ref{fig:results}. The dotted lines are the direct computation of the ideal quantum circuit in Fig.~\ref{fig:circuit}. The solid lines are the experimental results (given here for reference). The dashed lines are the simulation results with only one error source activated. 

The large difference between the ideal and real cases for the expression $\EP{3}$, seen in panels (c) of Figs.~\ref{fig:resultsPrep}, \ref{fig:resultsFeed} and \ref{fig:resultsMeas}, can be explained by the use of the imperfect backward trajectories and the divergence properties of $D$. We remind that the computation of  $\EP{3}$ is based on the comparison between the state before the forward evolution and the state after the backward evolution, for the two branches. One important feature of the backward protocol is that it can populate states that cannot be reached by the forward protocol in the same branch. Moreover, the relative entropy $D$, or divergence, comparing two probability distributions, is very sensitive to minor changes in the underlying distributions when their supports are different, \ie when there are states populated in one distribution but not in the other. Consequently, since the backward protocol in the presence of experimental imperfections populates states which don't appear in the forward protocol, the divergence based on the backward states becomes much more sensitive to these imperfections than any other expression of $\EP{}$.

\section{Minor experimental imperfections}\label{app:minor}

Besides the four major sources of system imperfections listed in the previous Section, there are several minor effects influencing the system evolution. We consider here the limited purity of the initial cavity state for the TPEM protocol, the atom and cavity relaxations and the possible presence of a second undetected atom in the main atom sample. By simulating the experimental sequence, we have found that the influence of these errors on the measured entropy production $\EP{}$ is relatively small, but still noticeable. Below we give the typical values of the corresponding errors.

The preparation of the cavity vacuum state $\ket{n=0}$ is exact. The preparation of the one-photon state by means of the resonant injection of a photon by a single excited atom results in the idle population of $0.08$ in $\ket{0}$ and $0.16$ in $\ket{2}$. The former is mainly due to the limited injection efficiency, while the latter originates from the possible presence of a second atom in the resonant atomic sample. The preparation of the two- and three-photon states are based on the photon-number measurements of a coherent state in the cavity~\cite{Guerlin07}. Being limited in time, it results in the non-zero populations in the two neighbouring states: for the two-photon preparation ($\ket{2}$) the cavity has a residual probability of $0.15$ to still contain one photon ($\ket{1}$) and $0.10$ to contain three photons ($\ket{3}$). For the three-photon state, $\ket{3}$, we get the residual populations in $\ket{2}$ and $\ket{4}$ equal to $0.17$ and $0.10$, respectively.

The lifetime of the circular Rydberg states used in the present experiment (principal quantum numbers $49$, $50$ and $51$) is of the order of $30$~ms. The cavity lifetime at a $1.5$-K temperature of the cryostat is $25$~ms. The duration of the main experimental sequence from the initial state preparation to the atom detection is less than $1$~ms. This limits the relaxation probability in the system to less than $0.04$. The atom and cavity relaxations are represented in the simulations with the corresponding master equations~\cite{HarocheBook, Peaudecerf13}.

The number $\na$ of atoms present in an atomic sample is random obeying the Poisson probability distribution $\Pa(\na)$. In the current work we have set the average atom number to $\bar{n}_\text{a}=0.22$ by adjusting the efficiency of the Rydberg state excitation of ground state atoms. The overall detection efficiency (\ie probability to detect an atom) is $\etad = 0.5$. For the data analysis we select trajectories with exactly one detected atom. The conditional probability to have two atoms in a trajectory containing only one detected atom is 
\begin{equation}
	P(\text{2}|\text{1\,detected}) = \frac{2\etad(1-\etad)\Pa(2)}{\etad \Pa(1) + 2\etad(1-\etad)\Pa(2)}.
	\label{eq:P2atoms}
\end{equation}
In our case, $P(\text{2}|\text{1\,detected}) = 0.10$.


\begin{thebibliography}{10}

\bibitem{Bejanbook2016}
A. Bejan, 
Advanced Engineering Thermodynamics,
4th edition (John Wiley \& Sons, Inc., New Jersey, 2016).

\bibitem{Esposito2009}
M. Esposito,U. Harbola,and S. Mukamel,
Nonequilibrium fluctuations, fluctuation theorems, and counting statistics in quantum systems,
Rev. Mod. Phys. \textbf{81}, 1665 (2009).

\bibitem{Campisi2011}
M. Campisi, P. H\"{a}nggi, and P. Talkner, 
Colloquium: Quantum fluctuation relations: Foundations and applications,
Rev. Mod. Phys. \textbf{83}, 771 (2011).

\bibitem{Binderbook2018}
F. Binder, L. A. Correa, C. Gogolin, J. Anders, and G. Adesso,
Thermodynamics in the Quantum Regime: Fundamental aspects and New Directions (Springer Nature Switzerland, 2018).

\bibitem{Landi2020}
G. T. Landi and M. Paternostro, 
Irreversible entropy production, from quantum to classical, 
arXiv:2009.07668v2 [quant-ph] (2020).

\bibitem{Elouard2017}
C. Elouard, D. A. Herrera-Martí, M. Clusel and Alexia Auffèves,
The role of quantum measurement in stochastic thermodynamics,
npj Quantum Inf. \textbf{3}, 9 (2017).

\bibitem{Maxwellbook1975}
J. C. Maxwell,
\emph{Theory of Heat}
(Longmans, Green, and Co., London, 1875), p. 328-329.

\bibitem{Rex2017}
A. Rex,
\textit{Maxwell's Demon --- A Historical Review},
Entropy \textbf{19}, 240 (2017).

\bibitem{Leff1990}
H. S. Leff and A. F. Rex (eds), 
\textit{Maxwell's Demon: Entropy, Information, Computing} (Princeton Univ. Press, 1990).

\bibitem{Batalhao2015}
T. B. Batalh\~ao, A. M. Souza, R. S. Sarthour, I. S. Oliveira, M. Paternostro, E. Lutz, and R. M. Serra, 
Irreversibility and the Arrow of Time in a Quenched Quantum System, 
Phys. Rev. Lett. \textbf{115}, 190601 (2015).

\bibitem{Brunelli2018}
M. Brunelli, L. Fusco, R. Landig, W. Wieczorek, J. Hoelscher-Obermaier, G. Landi, F.L. Semi\~ao, A. Ferraro, N. Kiesel, T. Donner, G. De Chiara, and M. Paternostro, 
Experimental Determination of Irreversible Entropy Production in out-of-Equilibrium Mesoscopic Quantum Systems, 
Phys. Rev. Lett. \textbf{121}, 160604 (2018).

\bibitem{Camati2016}
P. A. Camati, J. P. S. Peterson, T. B. Batalh\~ao, K. Micadei, A. M. Souza, R. S. Sarthour, I. S. Oliveira,
and R. M. Serra, 
Experimental Rectification of Entropy Production by Maxwell's Demon in a Quantum System, 
Phys. Rev. Lett. \textbf{117}, 240502 (2016).

\bibitem{Masuyama2018}
Y. Masuyama, K. Funo, Y. Murashita, A. Noguchi, S. Kono, Y. Tabuchi, R. Yamazaki, M. Ueda, and Y. Nakamura, Information-to-work conversion by Maxwell's demon in a superconducting
circuit quantum electrodynamical system, 
Nat. Comm. \textbf{9}, 1291 (2018).

\bibitem{Koski2015}
J. V. Koski, A. Kutvonen, I. M. Khaymovich, T. Ala-Nissila, and J. P. Pekola, 
On-Chip Maxwell's Demon as an Information-Powered Refrigerator,
Phys. Rev. Lett. \textbf{115}, 260602 (2015).

\bibitem{Luis2020}
B.-L. Najera-Santos, P. A. Camati, V. M\'etillon, M. Brune, J.-M. Raimond, A. Auff\`eves, and I. Dotsenko,
Autonomous Maxwell’s demon in a cavity QED system,
Phys. Rev. Research \textbf{2}, 032025(R) (2020).

\bibitem{An2015}
 S. An, J.-N. Zhang, M. Um, D. Lv, Y. Lu, J. Zhang, Z.-Q. Yin, H. T. Quan, and K. Kim,
Experimental test of the quantum Jarzynski equality with a trapped-ion system, 
Nat. Physics \textbf{11}, 193-199 (2015).

\bibitem{Huber2008}
G. Huber, F. Schmidt-Kaler, S. Deffner, and E. Lutz,
Employing trapped cold ions to verify the quantum Jarzynzki equality,
Phys. Rev. Lett. \textbf{101}, 070403 (2008).

\bibitem{Batalhao2014}
T. Batalhao, A. M. Souza, L. Mazzola, R. Auccaise, R. S. Sarthour, I. S. Oliveira, J. Goold, G. De Chiara, M. Paternostro, and R. M. Serra,
Experimental reconstruction of work distribution and verification of fluctuation relations at the full quantum level, Phys. Rev. Lett. \textbf{113}, 140601 (2014).

\bibitem{Heyl2012}
M. Heyl and S. Kehrein,
Crooks relation in optical spectra: Universality in work distributions for weak local quenches,
Phys. Rev. Lett. \textbf{108}, 190601 (2012).

\bibitem{Dorner2013}
R. Dorner, S. R. Clark, L. Heaney, R. Fazio, J. Goold, and V. Vedral, 
Extracting quantum work statistics and fluctuation theorems by single qubit interferometry,
Phys. Rev. Lett. \textbf{110}, 230601 (2013).

\bibitem{Mazzola2013}
L. Mazzola, G. D. Chiara, and M. Paternostro,
Measuring the characteristic function of the work distribution,
Phys. Rev. Lett. \textbf{110}, 230602 (2013).

\bibitem{Vedral2002}
V. Vedral,
The role of relative entropy in quantum information theory, 
Rev. Mod. Phys. \textbf{74}, 197 (2002).

\bibitem{Maruyama2009}
K. Maruyama, F. Nori, and V. Vedral,
Colloquium: The physics of Maxwell's demon and information,
Rev. Mod. Phys. \textbf{81}, 1 (2009).

\bibitem{Sagawa2009}
T. Sagawa and M. Ueda, 
Minimal Energy Cost for Thermodynamic Information Processing: Measurement and Information
Erasure,
Phys. Rev. Lett. \textbf{102}, 250602 (2009).

\bibitem{Sagawa2008}
T. Sagawa and M. Ueda, 
Second Law of Thermodynamics with Discrete Quantum Feedback Control,
Phys. Rev. Lett. \textbf{100}, 080403 (2008).

\bibitem{Deffner2011}
S. Deffner and E. Lutz, 
Nonequilibrium Entropy Production for Open Quantum Systems,
Phys. Rev. Lett. \textbf{107}, 140404 (2011).

\bibitem{Coverbook2006}
T. M. Cover and J. A. Thomas, 
Elements of Information Theory,
2nd ed. (John Wiley \& Sons, Inc., New Jersey, 2006).

\bibitem{Camati2021}
P. A. Camati, Z. Tan, G. Cœuret Cauquil, M. Brune, J.-M. Raimond, A. Auff\`eves, and I. Dotsenko,
in preparation.

\bibitem{Guerlin07}
C. Guerlin, J. Bernu, S. Del\'eglise, C. Sayrin, S. Gleyzes, S. Kuhr, M. Brune, J.-M. Raimond, and S. Haroche,
Progressive field-state collapse and quantum non-demolition photon counting,
Nature (London) \textbf{448}, 889 (2007).

\bibitem{Metillon19}
V. M\'etillon, S. Gerlichcorrections, M. Brune, J.M. Raimond, P. Rouchon, and I. Dotsenko,
Benchmarking maximum-likelihood state estimation with an entangled two-cavity state,
Phys. Rev. Lett. \textbf{123}, 060404 (2019).


\bibitem{HarocheBook}
S.~Haroche and J.M.~Raimond, Exploring the Quantum: atoms, cavities and photons, Oxford University Press, Oxford (2006).

\bibitem{Peaudecerf13}
B. Peaudecerf, C. Sayrin, X. Zhou, T. Rybarczyk, S. Gleyzes, I. Dotsenko, J. M. Raimond, M. Brune, and S. Haroche,
Quantum feedback experiments stabilizing Fock states of light in a cavity,
Phys. Rev. A \textbf{87}, 042320 (2013).





\end{thebibliography}
\end{document}